\errorstopmode
\input amssym.def
\input amssym.tex


\magnification=\magstephalf
\hsize=14.0 true cm
\vsize=19 true cm
\hoffset=1.0 true cm
\voffset=2.0 true cm

\abovedisplayskip=12pt plus 3pt minus 3pt
\belowdisplayskip=12pt plus 3pt minus 3pt
\parindent=1.0em


\font\sixrm=cmr6
\font\eightrm=cmr8
\font\ninerm=cmr9

\font\sixi=cmmi6
\font\eighti=cmmi8
\font\ninei=cmmi9

\font\sixsy=cmsy6
\font\eightsy=cmsy8
\font\ninesy=cmsy9

\font\sixbf=cmbx6
\font\eightbf=cmbx8
\font\ninebf=cmbx9

\font\eightit=cmti8
\font\nineit=cmti9

\font\eightsl=cmsl8
\font\ninesl=cmsl9

\font\sixss=cmss8 at 8 true pt
\font\sevenss=cmss9 at 9 true pt
\font\eightss=cmss8
\font\niness=cmss9
\font\tenss=cmss10

\font\sixmib=cmmib6
\font\sevenmib=cmmib7
\font\eightmib=cmmib8
\font\ninemib=cmmib9
\font\tenmib=cmmib10

 at 12 true pt
 at 12 true pt
\font\bigrm=cmr10 at 12 true pt
 at 12 true pt
 at 12 true pt

 at 16 true pt
 at 16 true pt
\font\Bigrm=cmr12 at 16 true pt
 at 16 true pt
 at 16 true pt

\catcode`@=11
\newfam\ssfam
\newfam\mibfam

\def\tenpoint{\def\rm{\fam0\tenrm}%
    \textfont0=\tenrm \scriptfont0=\sevenrm \scriptscriptfont0=\fiverm
    \textfont1=\teni  \scriptfont1=\seveni  \scriptscriptfont1=\fivei
    \textfont2=\tensy \scriptfont2=\sevensy \scriptscriptfont2=\fivesy
    \textfont3=\tenex \scriptfont3=\tenex   \scriptscriptfont3=\tenex
    \textfont\itfam=\tenit                  \def\it{\fam\itfam\tenit}%
    \textfont\slfam=\tensl                  \def\sl{\fam\slfam\tensl}%
    \textfont\bffam=\tenbf \scriptfont\bffam=\sevenbf
                           \scriptscriptfont\bffam=\fivebf
                           \def\bf{\fam\bffam\tenbf}%
    \textfont\ssfam=\tenss \scriptfont\ssfam=\sevenss
                           \scriptscriptfont\ssfam=\sevenss
                           \def\ss{\fam\ssfam\tenss}%
    \textfont\mibfam=\tenmib \scriptfont\mibfam=\sevenmib
                             \scriptscriptfont\mibfam=\sevenmib
                             \def\mib{\fam\mibfam\tenmib}%
    \normalbaselineskip=13pt
    \setbox\strutbox=\hbox{\vrule height8.5pt depth3.5pt width0pt}%
    \let\big=\tenbig
    \normalbaselines\rm}

\def\ninepoint{\def\rm{\fam0\ninerm}%
    \textfont0=\ninerm      \scriptfont0=\sixrm
                            \scriptscriptfont0=\fiverm
    \textfont1=\ninei       \scriptfont1=\sixi
                            \scriptscriptfont1=\fivei
    \textfont2=\ninesy      \scriptfont2=\sixsy
                            \scriptscriptfont2=\fivesy
    \textfont3=\tenex       \scriptfont3=\tenex
                            \scriptscriptfont3=\tenex
    \textfont\itfam=\nineit \def\it{\fam\itfam\nineit}%
    \textfont\slfam=\ninesl \def\sl{\fam\slfam\ninesl}%
    \textfont\bffam=\ninebf \scriptfont\bffam=\sixbf
                            \scriptscriptfont\bffam=\fivebf
                            \def\bf{\fam\bffam\ninebf}%
    \textfont\ssfam=\niness \scriptfont\ssfam=\sixss
                            \scriptscriptfont\ssfam=\sixss
                            \def\ss{\fam\ssfam\niness}%
    \textfont\mibfam=\ninemib \scriptfont\mibfam=\sixmib
                            \scriptscriptfont\mibfam=\sixmib
                            \def\mib{\fam\mibfam\ninemib}%
    \normalbaselineskip=12pt
    \setbox\strutbox=\hbox{\vrule height8.0pt depth3.0pt width0pt}%
    \let\big=\ninebig
    \normalbaselines\rm}

\def\eightpoint{\def\rm{\fam0\eightrm}%
    \textfont0=\eightrm      \scriptfont0=\sixrm
                             \scriptscriptfont0=\fiverm
    \textfont1=\eighti       \scriptfont1=\sixi
                             \scriptscriptfont1=\fivei
    \textfont2=\eightsy      \scriptfont2=\sixsy
                             \scriptscriptfont2=\fivesy
    \textfont3=\tenex        \scriptfont3=\tenex
                             \scriptscriptfont3=\tenex
    \textfont\itfam=\eightit \def\it{\fam\itfam\eightit}%
    \textfont\slfam=\eightsl \def\sl{\fam\slfam\eightsl}%
    \textfont\bffam=\eightbf \scriptfont\bffam=\sixbf
                             \scriptscriptfont\bffam=\fivebf
                             \def\bf{\fam\bffam\eightbf}%
    \textfont\ssfam=\eightss \scriptfont\ssfam=\sixss
                             \scriptscriptfont\ssfam=\sixss
                             \def\ss{\fam\ssfam\eightss}%
    \textfont\mibfam=\eightmib \scriptfont\mibfam=\sixmib
                             \scriptscriptfont\mibfam=\sixmib
                             \def\mib{\fam\mibfam\eightmib}%
    \normalbaselineskip=10pt
    \setbox\strutbox=\hbox{\vrule height7.0pt depth2.0pt width0pt}%
    \let\big=\eightbig
    \normalbaselines\rm}

\def\tenbig#1{{\hbox{$\left#1\vbox to8.5pt{}\right.\n@space$}}}
\def\ninebig#1{{\hbox{$\textfont0=\tenrm\textfont2=\tensy
                       \left#1\vbox to7.25pt{}\right.\n@space$}}}
\def\eightbig#1{{\hbox{$\textfont0=\ninerm\textfont2=\ninesy
                       \left#1\vbox to6.5pt{}\right.\n@space$}}}

\font\sectionfont=cmbx10
\font\subsectionfont=cmti10

\def\figurecaptionfont{\ninepoint}
\def\tablecaptionfont{\ninepoint}
\def\footnotefont{\eightpoint}


\newcount\equationno
\newcount\bibitemno
\newcount\figureno
\newcount\tableno

\equationno=0
\bibitemno=0
\figureno=0
\tableno=0


\footline={\ifnum\pageno=0{\hfil}\else
{\hss\rm\the\pageno\hss}\fi}


\def\section #1. #2 \par
{\vskip0pt plus .10\vsize\penalty-100 \vskip0pt plus-.10\vsize
\vskip 1.6 true cm plus 0.2 true cm minus 0.2 true cm
\global\def\equationlabel{#1}
\global\equationno=0
\leftline{\sectionfont #1. #2}\par
\immediate\write\terminal{Section #1. #2}
\vskip 0.7 true cm plus 0.1 true cm minus 0.1 true cm
\noindent}


\def\subsection #1 \par
{\vskip0pt plus 0.8 true cm\penalty-50 \vskip0pt plus-0.8 true cm
\vskip2.5ex plus 0.1ex minus 0.1ex
\leftline{\subsectionfont #1}\par
\immediate\write\terminal{Subsection #1}
\vskip1.0ex plus 0.1ex minus 0.1ex
\noindent}


\def\appendix #1. #2 \par
{\vskip0pt plus .20\vsize\penalty-100 \vskip0pt plus-.20\vsize
\vskip 1.6 true cm plus 0.2 true cm minus 0.2 true cm
\global\def\equationlabel{\hbox{\rm#1}}
\global\equationno=0
\leftline{\sectionfont Appendix #1. #2}\par
\immediate\write\terminal{Appendix #1. #2}
\vskip 0.7 true cm plus 0.1 true cm minus 0.1 true cm
\noindent}



\def\equation#1{$$\displaylines{\qquad #1}$$}
\def\enum{\global\advance\equationno by 1
\hfill\llap{{\rm(\equationlabel.\the\equationno)}}}
\def\noenum{\hfill}
\def\next#1{\cr\noalign{\vskip#1}\qquad}


\def\ifundefined#1{\expandafter\ifx\csname#1\endcsname\relax}

\def\ref#1{\ifundefined{#1}?\immediate\write\terminal{unknown reference
on page \the\pageno}\else\csname#1\endcsname\fi}

\newwrite\terminal
\newwrite\bibitemlist

\def\bibitem#1#2\par{\global\advance\bibitemno by 1
\immediate\write\bibitemlist{\string\def
\expandafter\string\csname#1\endcsname
{\the\bibitemno}}
\item{[\the\bibitemno]}#2\par}

\def\beginbibliography{
\vskip0pt plus .15\vsize\penalty-100 \vskip0pt plus-.15\vsize
\vskip 1.2 true cm plus 0.2 true cm minus 0.2 true cm
\leftline{\sectionfont References}\par
\immediate\write\terminal{References}
\immediate\openout\bibitemlist=biblist
\frenchspacing\parindent=1.8em
\vskip 0.5 true cm plus 0.1 true cm minus 0.1 true cm}

\def\endbibliography{
\immediate\closeout\bibitemlist
\nonfrenchspacing\parindent=1.0em}


\def\figurecaption#1{\global\advance\figureno by 1
\narrower\figurecaptionfont
Fig.~\the\figureno. #1}

\def\tablecaption#1{\global\advance\tableno by 1
\vbox to 0.25 true cm { }
\centerline{\tablecaptionfont%
Table~\the\tableno. #1}
\vskip-0.4 true cm}

\def\thicktablerule{\hrule height1pt}
\def\thintablerule{\hrule height0.4pt}

\tenpoint

\immediate\openin\bibitemlist=biblist
\ifeof\bibitemlist\immediate\closein\bibitemlist
\else\immediate\closein\bibitemlist
\input biblist \fi


\def\thismonth{\ifcase\month\or
January\or February\or March\or April\or May\or June\or
July\or August\or September\or October\or November\or December\fi}

\input epsf
\epsfclipon



\def\rme{{\rm e}}
\def\rmO{{\rm O}}



\def\proof{\noindent{\sl Proof:}\kern0.6em}

\def\frac#1#2{\hbox{$#1\over#2$}}
\def\dual{\mathstrut^*\kern-0.1em}

\def\lvec#1{\setbox0=\hbox{$#1$}
    \setbox1=\hbox{$\scriptstyle\leftarrow$}
    #1\kern-\wd0\smash{
    \raise\ht0\hbox{$\raise1pt\hbox{$\scriptstyle\leftarrow$}$}}
    \kern-\wd1\kern\wd0}
\def\rvec#1{\setbox0=\hbox{$#1$}
    \setbox1=\hbox{$\scriptstyle\rightarrow$}
    #1\kern-\wd0\smash{
    \raise\ht0\hbox{$\raise1pt\hbox{$\scriptstyle\rightarrow$}$}}
    \kern-\wd1\kern\wd0}
\def\slash#1{\setbox0=\hbox{$#1$}\setbox1=\hbox{$\kern1pt/$}
    #1\kern-\wd0\kern1pt/\kern-\wd1\kern\wd0}


\def\nabstar#1{{\nabla\kern0.5pt\smash{\raise 4.5pt\hbox{$\ast$}}
               \kern-5.5pt_{#1}}}
\def\drv#1{{\partial_{#1}}}
\def\drvstar#1{{\partial\kern0.5pt\smash{\raise 4.5pt\hbox{$\ast$}}
               \kern-6.0pt_{#1}}}

\def\ldrvstar#1{{\lvec{\,\partial}\kern-0.5pt\smash{\raise 4.5pt\hbox{$\ast$}}
               \kern-5.0pt_{#1}}}




\def\rbar{\bar{r}}


\def\dirac#1{\gamma_{#1}}
\def\diracstar#1#2{
    \setbox0=\hbox{$\gamma$}\setbox1=\hbox{$\gamma_{#1}$}
    \gamma_{#1}\kern-\wd1\kern\wd0
    \smash{\raise4.5pt\hbox{$\scriptstyle#2$}}}


\def\tr{{\rm tr}}

\def\Ad{{\rm Ad}\kern0.1em}


\def\fpp{f_{\rm PP}}
\def\fap{f_{\rm AP}}
\def\fvv{f_{\rm VV}}


\def\mpi{M_{\pi}}
\def\Mpi{\mpi}

\def\Fpi{F_{\pi}}

\def\Rpi{R_{\pi}}
\def\Mps{M_{\rm PS}}
\def\Fps{F_{\rm PS}}
\def\Gps{G_{\rm PS}}
\def\Rps{R_{\rm PS}}
\def\Mv{M_{\rm V}}

\def\mbar{\overline{m\kern-0.07em}\kern0.07em}
\def\meff{m_{\rm eff}}
\def\Meff{M_{\rm eff}}

\def\Geff{G_{\rm eff}}
\def\Peff{P_{\rm eff}}

\def\msea{m_{\rm sea}}
\def\mval{m_{\rm val}}
\def\ZA{Z_{\rm A}}
\def\ZP{Z_{\rm P}}
\def\ZX{Z_{\rm X}}
\def\mubar{\bar{\mu}}
\def\ksea{\kappa_{\rm sea}}


\def\csw{c_{\rm sw}}
\def\cA{c_{\rm A}}

\def\bXsea{\bar{b}_{\rm X}}
\def\bXval{\tilde{b}_{\rm X}}
\def\Ncfg{N_{\rm cfg}}

\def\Ntrj{N_{\rm trj}}
\def\Nsep{N_{\rm sep}}
\def\Ncg{N_{\rm CG}}
\def\Ngcr{N_{\rm GCR}}

\def\tauint{\tau_{\rm int}}

\def\abar{\bar{a}}
\def\bbar{\bar{b}}
\def\fbar{\bar{f}}
\def\gbar{\bar{g}}
\def\eps{\epsilon}
\def\del{\delta}
%
\rightline{CERN-PH-TH/2006-261}

\vskip1.0cm 
\centerline{\Bigrm
QCD with light Wilson quarks on fine lattices (II):
}
\vskip 0.3 true cm
\centerline{\Bigrm
DD-HMC simulations and data analysis
}
\vskip 0.6cm
\centerline{\bigrm L.~Del Debbio$^{1}$, L.~Giusti$^{2,}$%
\footnote{$^{\ast}$}{\footnotefont%
On leave from
Centre de Physique Th\'eorique, CNRS Luminy, F-13288 Marseille, France},
M.~L\"uscher$^2$, R.~Petronzio$^3$, N.~Tantalo$^{3,4}$}
\vskip2.0ex
\centerline{$^1\hskip-3pt$ \it
SUPA, School of Physics, University of Edinburgh, Edinburgh EH9 3JZ, UK}
\vskip1.0ex
\centerline{$^2\hskip-3pt$ \it
CERN, Physics Department, TH Division, CH-1211 Geneva 23, Switzerland}
\vskip1.0ex
\centerline{$^3\hskip-3pt$ \it
Universit\`a di Roma ``Tor Vergata'' and INFN sezione ``Tor Vergata'',}
\centerline{\it Via della Ricerca Scientifica 1,  I-00133 Rome, Italy}
\vskip1.0ex
\centerline{$^4\hskip-3pt$ \it
Centro Enrico Fermi, Via Panisperna 89 A, I-00184 Rome, Italy}
\vskip 0.8 true cm
\thintablerule
\vskip 2.0ex
\ninepoint
\leftline{\bf Abstract}
\vskip 1.0ex\noindent
In this second report on our recent numerical simulations
of two-flavour QCD, we provide further technical details on 
the simulations and describe the 
methods we used to extract the meson masses and 
decay constants from the generated ensembles of gauge fields.
Among the topics covered are the
choice of the DD-HMC parameters,
the issue of stability, autocorrelations and the statistical 
error analysis.
Extensive data tables are included as well as 
a short discussion of the quark-mass dependence 
in partially quenched QCD, supplementing
the physics analysis that was presented in the first paper 
in this series.
\vskip 2.0ex
\thintablerule

\tenpoint

\vskip-0.2cm

\section 1. Introduction

Lattice QCD with Wilson quarks [\ref{Wilson}] has seen 
important algorithmic developments in the last few years
[\ref{Hasenbusch}--\ref{UrbachEtAl}].
As a consequence, 
a large range of lattice spacings,
lattice volumes and quark masses can now be explored,
using numerical simulations, 
thus providing new physics opportunities and
a greater lever arm for the extrapolations to the continuum 
and the chiral limit.
Our recent work [\ref{I}] was the first to fully profit from
the technical breakthrough and several other projects, 
simulating QCD with two [\ref{GoeckelerEtAl},\ref{MeyerEtAl}] and 
three [\ref{IshikawaEtAl}--\ref{UkawaEtAl}] flavours 
of light Wilson quarks, or with two flavours and 
a twisted mass term [\ref{JansenUrbach}], 
are currently underway, all heavily
depending on the new generation of algorithms.

The present paper is the second in a series of two papers
devoted to the study of two-flavour QCD at small
quark masses and lattice spacings. In the first paper [\ref{I}],
the focus was on the physics results, while here  
we give a fairly detailed technical account of 
the simulations that we have performed. 

Perhaps the most important items that we discuss
are the stability of the simulations (sect.~3)
and the pattern of autocorrelation times observed 
in our runs (sect.~4).
We also describe, in sect.~5, 
the methods that we used to extract the meson masses 
and decay constants from the 
generated ensembles of gauge fields (extensive
data tables are included in appendix C).
The paper ends with an addendum to the first paper, where 
we briefly discuss the quark-mass dependence of various quantities
in partially quenched QCD with $2+1$ flavours of quarks.

\section 2. Simulation parameters

We consider the Wilson formulation of lattice QCD, optionally
O($a$)-improved, with a doublet of mass-degenerate sea quarks.
The notation and normalization conventions adopted in this paper
coincide with those already used in our previous paper [\ref{I}].
In particular, the parameters of the lattice theory are the inverse
bare coupling $\beta$, the sea-quark hopping parameter $\ksea$
and the coefficient $\csw$ of the Sheikoleslami-Wohlert improvement
term [\ref{SW},\ref{OaImp}].

All simulations reported here
were performed using the DD-HMC
simulation algorithm [\ref{SchwarzIII}].
As suggested by the name,
the algorithm combines domain-decomposition ideas with 
the HMC algorithm [\ref{HMC}]. 
More precisely, by dividing the lattice
into non-overlapping rectangular blocks,
a natural separation of the high-frequency 
from the low-frequency modes of the fields is achieved.
Following
Sexton and Weingarten [\ref{SextonWeingarten}],
the different modes are then evolved
using different molecular-dynamics step sizes, 
which results in a significant acceleration of the simulation.

\topinsert
\newdimen\digitwidth
\setbox0=\hbox{\rm 0}
\digitwidth=\wd0
\catcode`@=\active
\def@{\kern\digitwidth}
\tablecaption{Lattice parameters and simulation statistics} 
\vskip-0.5ex
$$\vbox{\settabs\+&%
                  xxxxxx&x&
                  xxxxxxxxx&xx&
                  xxxxx&xx&
                  xxxxxxxx&xx&
                  xxxxxxxx&x&
                  xxxxxxx&x&
                  xxxxxx&x&
                  xxxxxx&\cr
\thicktablerule
\vskip1.0ex
                \+& \hfill Run\hfill
                 && \hfill Lattice\hfill
                 && \hfill $\beta$\hfill
                 && \hfill $\csw$\hfill
                 && \hfill $\ksea$\hfill
                 && \hfill $\Ntrj$\hfill
                 && \hfill $\Nsep$\hfill
                 && \hfill $\Ncfg$\hfill
                 &\cr
\vskip1.0ex
\thintablerule
\vskip1.2ex
  \+& \hskip2.5ex $A_{1a}$\hfill
  &&  \hfill $32\times24^3$\hfill
  &&  \hfill $5.6$\hfill 
  &&  \hfill $0$\hfill
  &&  \hfill $0.15750$\hfill
  &&  \hfill $@6300$\hfill
  &&  \hfill $100$\hfill
  &&  \hfill $@64$\hfill
  &\cr
\vskip0.3ex
  \+& \hskip2.5ex $A_{1b}$\hfill
  &&  \hfill $$\hfill
  &&  \hfill $$\hfill 
  &&  \hfill $$\hfill
  &&  \hfill $0.15750$\hfill
  &&  \hfill $@5070$\hfill
  &&  \hfill $@30$\hfill
  &&  \hfill $169$\hfill
  &\cr
\vskip0.3ex
  \+& \hskip2.5ex $A_2$\hfill
  &&  \hfill $$\hfill
  &&  \hfill $$\hfill 
  &&  \hfill $$\hfill
  &&  \hfill $0.15800$\hfill
  &&  \hfill $10800$\hfill
  &&  \hfill $100$\hfill
  &&  \hfill $109$\hfill
  &\cr
\vskip0.3ex
  \+& \hskip2.5ex $A_{3a}$\hfill
  &&  \hfill $$\hfill
  &&  \hfill $$\hfill 
  &&  \hfill $$\hfill
  &&  \hfill $0.15825$\hfill
  &&  \hfill $@6100$\hfill
  &&  \hfill $100$\hfill
  &&  \hfill $@62$\hfill
  &\cr
\vskip0.3ex
  \+& \hskip2.5ex $A_{3b}$\hfill
  &&  \hfill $$\hfill
  &&  \hfill $$\hfill 
  &&  \hfill $$\hfill
  &&  \hfill $0.15825$\hfill
  &&  \hfill $@3800$\hfill
  &&  \hfill $100$\hfill
  &&  \hfill $@38$\hfill
  &\cr
\vskip0.3ex
  \+& \hskip2.5ex $A_{4}$\hfill
  &&  \hfill $$\hfill
  &&  \hfill $$\hfill 
  &&  \hfill $$\hfill
  &&  \hfill $0.15835$\hfill
  &&  \hfill $@4950$\hfill
  &&  \hfill $@50$\hfill
  &&  \hfill $100$\hfill
  &\cr
\vskip0.3ex
  \+& \hskip2.5ex $B_1$\hfill
  &&  \hfill $64\times32^3$\hfill
  &&  \hfill $5.8$\hfill 
  &&  \hfill $0$\hfill
  &&  \hfill $0.15410$\hfill
  &&  \hfill $@5050$\hfill
  &&  \hfill $@50$\hfill
  &&  \hfill $100$\hfill
  &\cr
\vskip0.3ex
  \+& \hskip2.5ex $B_2$\hfill
  &&  \hfill $$\hfill
  &&  \hfill $$\hfill 
  &&  \hfill $$\hfill
  &&  \hfill $0.15440$\hfill
  &&  \hfill $@5200$\hfill
  &&  \hfill $@50$\hfill
  &&  \hfill $101$\hfill
  &\cr
\vskip0.3ex
  \+& \hskip2.5ex $B_3$\hfill
  &&  \hfill $$\hfill
  &&  \hfill $$\hfill 
  &&  \hfill $$\hfill
  &&  \hfill $0.15455$\hfill
  &&  \hfill $@5150$\hfill
  &&  \hfill $@50$\hfill
  &&  \hfill $104$\hfill
  &\cr
\vskip0.3ex
  \+& \hskip2.5ex $B_4$\hfill
  &&  \hfill $$\hfill
  &&  \hfill $$\hfill 
  &&  \hfill $$\hfill
  &&  \hfill $0.15462$\hfill
  &&  \hfill $@5050$\hfill
  &&  \hfill $@50$\hfill
  &&  \hfill $102$\hfill
  &\cr
\vskip0.3ex
  \+& \hskip2.5ex $C_1$\hfill
  &&  \hfill $64\times24^3$\hfill
  &&  \hfill $5.6$\hfill 
  &&  \hfill $0$\hfill
  &&  \hfill $0.15800$\hfill
  &&  \hfill $@3450$\hfill
  &&  \hfill $@30$\hfill
  &&  \hfill $116$\hfill
  &\cr
\vskip0.3ex
  \+& \hskip2.5ex $D_1$\hfill
  &&  \hfill $48\times24^3$\hfill
  &&  \hfill $5.3$\hfill 
  &&  \hfill $1.90952$\hfill
  &&  \hfill $0.13550$\hfill
  &&  \hfill $@5150$\hfill
  &&  \hfill $@50$\hfill
  &&  \hfill $104$\hfill
  &\cr
\vskip0.3ex
  \+& \hskip2.5ex $D_2$\hfill
  &&  \hfill $$\hfill
  &&  \hfill $$\hfill 
  &&  \hfill $$\hfill
  &&  \hfill $0.13590$\hfill
  &&  \hfill $@5130$\hfill
  &&  \hfill $@30$\hfill
  &&  \hfill $171$\hfill
  &\cr
\vskip0.3ex
  \+& \hskip2.5ex $D_3$\hfill
  &&  \hfill $$\hfill
  &&  \hfill $$\hfill 
  &&  \hfill $$\hfill
  &&  \hfill $0.13610$\hfill
  &&  \hfill $@5040$\hfill
  &&  \hfill $@30$\hfill
  &&  \hfill $168$\hfill
  &\cr
\vskip0.3ex
  \+& \hskip2.5ex $D_4$\hfill
  &&  \hfill $$\hfill
  &&  \hfill $$\hfill 
  &&  \hfill $$\hfill
  &&  \hfill $0.13620$\hfill
  &&  \hfill $@5010$\hfill
  &&  \hfill $@30$\hfill
  &&  \hfill $168$\hfill
  &\cr
\vskip0.3ex
  \+& \hskip2.5ex $D_5$\hfill
  &&  \hfill $$\hfill
  &&  \hfill $$\hfill 
  &&  \hfill $$\hfill
  &&  \hfill $0.13625$\hfill
  &&  \hfill $@5040$\hfill
  &&  \hfill $@30$\hfill
  &&  \hfill $169$\hfill
  &\cr
\vskip0.3ex
  \+& \hskip2.5ex $E_1$\hfill
  &&  \hfill $64\times32^3$\hfill
  &&  \hfill $5.3$\hfill 
  &&  \hfill $1.90952$\hfill
  &&  \hfill $0.13550$\hfill
  &&  \hfill $@5344$\hfill
  &&  \hfill $@32$\hfill
  &&  \hfill $168$\hfill
  &\cr
\vskip0.3ex
  \+& \hskip2.5ex $E_2$\hfill
  &&  \hfill $$\hfill
  &&  \hfill $$\hfill 
  &&  \hfill $$\hfill
  &&  \hfill $0.13590$\hfill
  &&  \hfill $@5024$\hfill
  &&  \hfill $@32$\hfill
  &&  \hfill $158$\hfill
  &\cr
\vskip0.3ex
  \+& \hskip2.5ex $E_3$\hfill
  &&  \hfill $$\hfill
  &&  \hfill $$\hfill 
  &&  \hfill $$\hfill
  &&  \hfill $0.13605$\hfill
  &&  \hfill $@5024$\hfill
  &&  \hfill $@32$\hfill
  &&  \hfill $158$\hfill
  &\cr
\vskip1.2ex
\thicktablerule
}
$$
\vskip-2.0ex
\endinsert

On a given lattice and at fixed coupling, 
the simulations progressed 
from the larger to the smaller quark masses, 
normally skipping $1500$ molecular-dynamics 
trajectories for thermalization.
The number $\Ntrj$ of trajectories generated after thermalization,
the separation $\Nsep$ (in numbers of trajectories) between 
successive saved field configurations and the number $\Ncfg$ of
saved fields are given in table~1.
Different runs at the same lattice parameters
(such as $A_{3a}$ and $A_{3b}$) are distinguished by a lower-case 
latin index.
In our previous paper [\ref{I}], only the runs 
$A_{1a}$, $A_2$, $A_{3a}$, $A_{3b}$, $B_1$--$B_4$ and $D_1$--$D_5$
were included in the physics analysis.
The other runs listed in table~1
merely serve, in sections $3$ and $4$, to clarify  
some technical issues.

\topinsert
\newdimen\digitwidth
\setbox0=\hbox{$0$}
\digitwidth=\wd0
\catcode`@=\active
\def@{\kern\digitwidth}
\tablecaption{DD-HMC parameters, acceptance rate and average 
solver iteration numbers} 
\vskip-1.0ex

$$\vbox{\settabs\+&%
                  xxxxxx&xx&
                  xxxxxxxxxxxxxxxx&x&
                  xxxxxxx&xx&
                  xxxxxxxx&xx&
                  xxxxxxxx&xx&
                  xxxxxxxx&xx\cr
                  
\thicktablerule
\vskip1.0ex
                \+& \hfill Run\hfill
                 && \hfill Block size\hfill
                 && \hfill $N_2$\hfill
                 && \hfill $P_{\rm acc}$\hfill
                 && \hfill $\langle\Ngcr\rangle$\hfill
                 && \hfill $\langle\Ncg\rangle$\hfill
                 &\cr
\vskip1.0ex
\thintablerule
\vskip1.5ex
  \+& \hskip2.0ex $A_{1a}$\hfill
  &&  \hfill $8\times6^2\times12$\hfill
  &&  \hfill $@5$\hfill 
  &&  \hfill $0.81^*$\hfill
  &&  \hfill $23$\hfill
  &&  \hfill $@73$\hfill 
  &\cr
\vskip0.3ex
  \+& \hskip2.0ex $A_{1b}$\hfill
  &&  \hfill $8\times6\times12^2$\hfill
  &&  \hfill $@5$\hfill 
  &&  \hfill $0.82\phantom{^*}$\hfill
  &&  \hfill $22$\hfill 
  &&  \hfill $@89$\hfill
  &\cr
\vskip0.3ex
  \+& \hskip2.0ex $A_{2}$\hfill
  &&  \hfill $8\times6^2\times12$\hfill
  &&  \hfill $@6$\hfill 
  &&  \hfill $0.79^*$\hfill
  &&  \hfill $39$\hfill 
  &&  \hfill $@74$\hfill
  &\cr
\vskip0.3ex
  \+& \hskip2.0ex $A_{3a}$\hfill
  &&  \hfill $$\hfill
  &&  \hfill $10$\hfill 
  &&  \hfill $0.89^*$\hfill
  &&  \hfill $54$\hfill 
  &&  \hfill $@75$\hfill
  &\cr
\vskip0.3ex
  \+& \hskip2.0ex $A_{3b}$\hfill
  &&  \hfill $$\hfill
  &&  \hfill $10$\hfill 
  &&  \hfill $0.86\phantom{^*}$\hfill
  &&  \hfill $54$\hfill 
  &&  \hfill $@75$\hfill
  &\cr
\vskip0.3ex
  \+& \hskip2.0ex $A_4$\hfill
  &&  \hfill $$\hfill
  &&  \hfill $16$\hfill 
  &&  \hfill $0.91\phantom{^*}$\hfill
  &&  \hfill $73$\hfill 
  &&  \hfill $@75$\hfill
  &\cr
\vskip0.3ex
  \+& \hskip2.0ex $B_{1}$\hfill
  &&  \hfill $8^3\times16$\hfill
  &&  \hfill $@8$\hfill 
  &&  \hfill $0.84\phantom{^*}$\hfill
  &&  \hfill $32$\hfill 
  &&  \hfill $@85$\hfill
  &\cr
\vskip0.3ex
  \+& \hskip2.0ex $B_{2}$\hfill
  &&  \hfill $$\hfill
  &&  \hfill $10$\hfill 
  &&  \hfill $0.89\phantom{^*}$\hfill
  &&  \hfill $52$\hfill 
  &&  \hfill $@87$\hfill
  &\cr
\vskip0.3ex
  \+& \hskip2.0ex $B_{3}$\hfill
  &&  \hfill $$\hfill
  &&  \hfill $12$\hfill 
  &&  \hfill $0.87\phantom{^*}$\hfill
  &&  \hfill $74$\hfill 
  &&  \hfill $@87$\hfill
  &\cr
\vskip0.3ex
  \+& \hskip2.0ex $B_{4}$\hfill
  &&  \hfill $$\hfill
  &&  \hfill $14$\hfill 
  &&  \hfill $0.92\phantom{^*}$\hfill
  &&  \hfill $90$\hfill 
  &&  \hfill $@88$\hfill
  &\cr
\vskip0.3ex
  \+& \hskip2.0ex $C_{1}$\hfill
  &&  \hfill $8\times6\times12^2$\hfill
  &&  \hfill $@7$\hfill 
  &&  \hfill $0.81\phantom{^*}$\hfill
  &&  \hfill $41$\hfill 
  &&  \hfill $@92$\hfill
  &\cr
\vskip0.3ex
  \+& \hskip2.0ex $D_{1}$\hfill
  &&  \hfill $6^2\times12^2$\hfill
  &&  \hfill $@7$\hfill 
  &&  \hfill $0.81\phantom{^*}$\hfill
  &&  \hfill $25$\hfill 
  &&  \hfill $120$\hfill
  &\cr
\vskip0.3ex
  \+& \hskip2.0ex $D_{2}$\hfill
  &&  \hfill $$\hfill
  &&  \hfill $@8$\hfill 
  &&  \hfill $0.80\phantom{^*}$\hfill
  &&  \hfill $41$\hfill 
  &&  \hfill $123$\hfill
  &\cr
\vskip0.3ex
  \+& \hskip2.0ex $D_{3}$\hfill
  &&  \hfill $$\hfill
  &&  \hfill $12$\hfill 
  &&  \hfill $0.87\phantom{^*}$\hfill
  &&  \hfill $58$\hfill 
  &&  \hfill $124$\hfill
  &\cr
\vskip0.3ex
  \+& \hskip2.0ex $D_{4}$\hfill
  &&  \hfill $$\hfill
  &&  \hfill $14$\hfill 
  &&  \hfill $0.87\phantom{^*}$\hfill
  &&  \hfill $73$\hfill 
  &&  \hfill $125$\hfill
  &\cr
\vskip0.3ex
  \+& \hskip2.0ex $D_{5}$\hfill
  &&  \hfill $$\hfill
  &&  \hfill $18$\hfill 
  &&  \hfill $0.89\phantom{^*}$\hfill
  &&  \hfill $87$\hfill 
  &&  \hfill $125$\hfill
  &\cr
\vskip0.3ex
  \+& \hskip2.0ex $E_{1}$\hfill
  &&  \hfill $8^4$\hfill
  &&  \hfill $@9$\hfill 
  &&  \hfill $0.80\phantom{^*}$\hfill
  &&  \hfill $25$\hfill 
  &&  \hfill $121$\hfill
  &\cr
\vskip0.3ex
  \+& \hskip2.0ex $E_{2}$\hfill
  &&  \hfill $$\hfill
  &&  \hfill $11$\hfill 
  &&  \hfill $0.84\phantom{^*}$\hfill
  &&  \hfill $41$\hfill 
  &&  \hfill $124$\hfill
  &\cr
\vskip0.3ex
  \+& \hskip2.0ex $E_{3}$\hfill
  &&  \hfill $$\hfill
  &&  \hfill $13$\hfill 
  &&  \hfill $0.83\phantom{^*}$\hfill
  &&  \hfill $53$\hfill 
  &&  \hfill $125$\hfill
  &\cr
\vskip1.5ex
\thicktablerule
\vskip1.5ex
\+&{\footnotefont$^*$ Transition probability includes trajectory replays}&\cr
}
$$
\vskip-2ex
\endinsert

The DD-HMC simulation algorithm was implemented following the lines of
ref.~[\ref{SchwarzIII}]. In particular, for the solution of 
the Dirac equation on the full lattice, the Schwarz-preconditioned
GCR solver described in ref.~[\ref{SchwarzII}] was used.
The so-called replay trick, however, was
switched off in the more recent simulations $A_{3b}$--$E_3$,
because trajectory replays would have been rare
and hardly worth the extra effort (see subsect.~3.3).

No attempt was made to tune the DD-HMC algorithm
and most of its parameters were actually
set to some fixed values, the same as the ones already
chosen in ref.~[\ref{SchwarzIII}].
Among these were the trajectory
length $\tau=0.5$, the integration step numbers $N_0=4$ and $N_1=5$
associated to the gauge and block fermion forces as well as
the admitted tolerances 
$(r_1,r_2,\tilde{r}_1,\tilde{r}_2)=(10^{-8},10^{-7},10^{-11},10^{-10})$
for the numerical solution of the Dirac equation on the 
blocks and the full lattice%
\kern1.5pt\footnote{$\dagger$}{\footnotefont%
The trajectory length $\tau$ and thus the 
integration step sizes $\tau/N_2$, etc., refer to 
a particular normalization of the kinetic term 
in the molecular-dynamics Hamiltonian.
Here the normalizations are the same as in
ref.~[\ref{SchwarzIII}], i.e.~the term is assumed to be equal to
$\frac{1}{2}(\Pi,\Pi)=\sum_{x,\mu}\tr\{\Pi(x,\mu)^{\dagger}\Pi(x,\mu)\}$,
where $\Pi(x,\mu)$ denotes the canonical momentum of the link variable
$U(x,\mu)$. 
}.
The parameters of the Schwarz-preconditioned GCR solver
were fixed to the values quoted
in ref.~[\ref{SchwarzII}], except for
the number $n_{\rm kv}$ of Krylov vectors generated 
before the GCR recursion is restarted, which was set to $32$
in run $D_5$ and to $24$ in all other runs. 

What remains to be specified are then 
the size of the blocks on which the 
algorithm operates and the integration step number $N_2$ 
associated to the block interaction term in the molecular-dynamics
Hamiltonian (see table~2). In practice the latter must be increased
as one moves to lighter quark masses in order to 
preserve a high acceptance rate $P_{\rm acc}$. 
The average number $\Ngcr$ 
of GCR solver iterations needed along the trajectories
also depends on $N_2$ (it decreases when $N_2$ goes up),
while the average number $\Ncg$ of conjugate-gradient iterations required
for the computation of the block terms in 
the molecular-dynamics equations is largely
determined by the block size.

With the chosen parameters, the reversibility of the molecular-dynamics
trajectories is guaranteed to high precision. In the tests
that we have performed, the average absolute deviation of the 
components of the link variables after a return trajectory was 
at most $3\times10^{-9}$, while in the case of the Hamiltonian
the observed differences were less than $4\times10^{-6}$. Deviations
larger than $10$ times the average occurred in less than $1\%$ of
the cases and never went beyond $100$ times the average.

\section 3. Spectral gap and stability issues

The Wilson--Dirac operator preserves chiral symmetry only
up to lattice effects and is therefore not rigorously protected
from having eigenvalues much smaller 
than the quark mass.
Exceptionally small eigenvalues do not invalidate the theory
but may lead to instabilities in numerical
simulations, depending, to some extent, on which simulation
algorithm is used.

\subsection 3.1 Spectral gap of the Dirac operator

In a previous dedicated study [\ref{Stability}], 
we computed the distribution of the spectral gap of the 
hermitian lattice Dirac operator on the lattices 
$A_1-A_4$, $B_1$, $B_2$, $C_1$ and $D_1$.
The distributions turned out to be well separated from the origin,
thus showing, a posteriori, that the simulations
were safe of exceptionally small eigenvalues and the 
associated instabilities.
Moreover, based on the observed 
scaling properties of the distributions on the $A$, $B$ and $C$
lattices,
we argued that this will always be so in the large-volume regime
of the unimproved Wilson theory.

\topinsert
\vbox{
\vskip0.0cm
\epsfxsize=5.6cm\hskip3.2cm\epsfbox{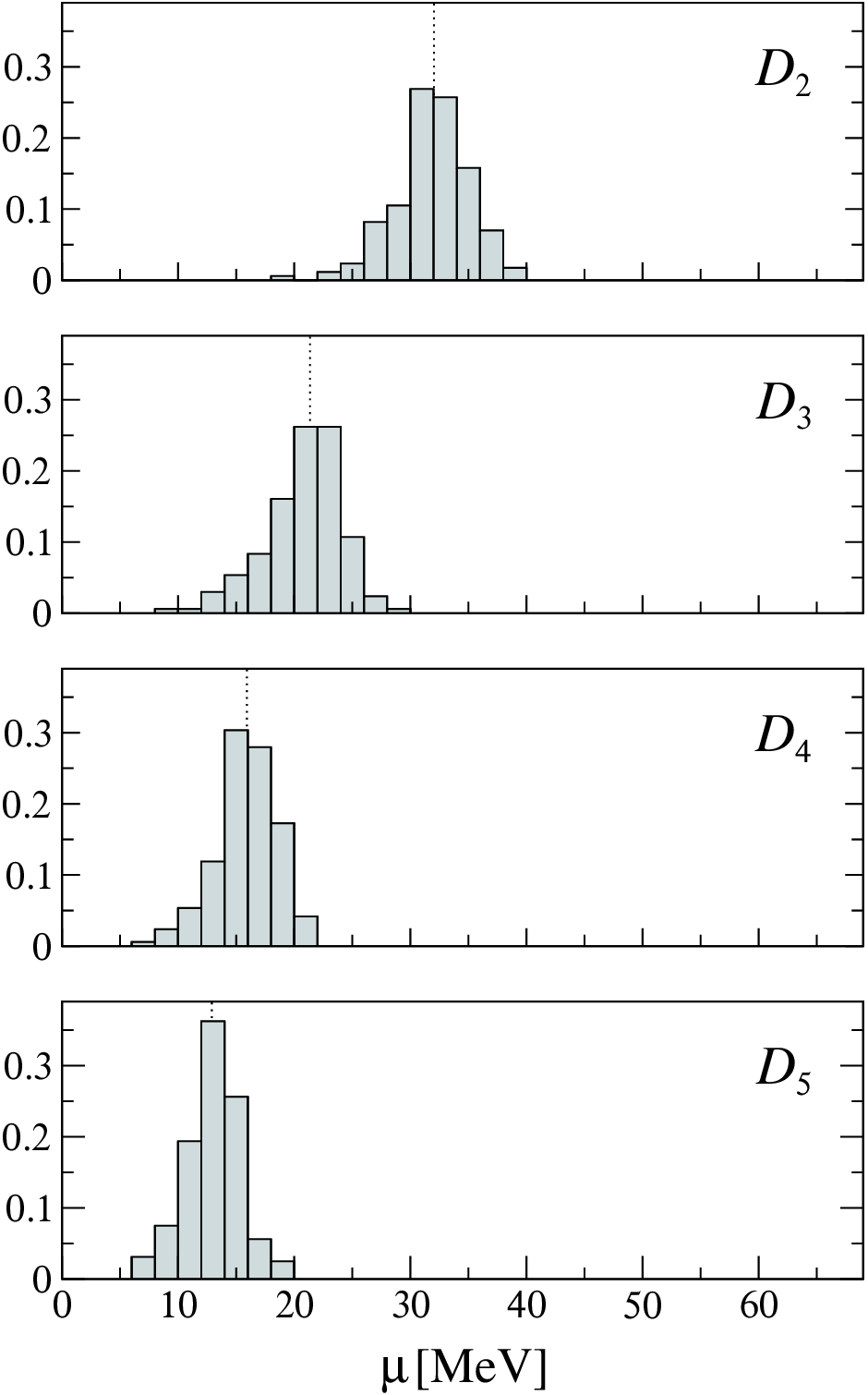}
\vskip0.3cm
\figurecaption{%
Normalized histograms of the (unrenormalized) 
spectral gap $\mu$ of the hermitian lattice
Dirac operator, as obtained in the runs $D_2-D_5$. 
The bin size is $2$ MeV
and the dotted vertical lines indicate the position of 
the median $\mubar$ of the distributions. The data were 
converted to physical units using $a=0.0784$ fm [\ref{I}].
}
\vskip0.0cm
}
\endinsert

The gap distributions have now also been computed on the 
lattices $B_3,B_4$, $D_2-D_5$, $E_2$ and $E_3$.
In the following, however, we focus on the improved theory,
because the results obtained on the $B$ lattices
are fully in line with the behaviour expected from 
our previous paper [\ref{Stability}].

At first sight, the gap distributions in the improved theory 
look similar to those in the unimproved theory
(see fig.~1). In particular, they are well separated from 
the origin, on all lattices that we 
have simulated, and the median of the distributions 
again turns out to be a practically 
linear function of the sea-quark mass (fig.~2).

\topinsert
\vbox{
\vskip0.0cm
\epsfysize=5.0cm\hskip2.2cm\epsfbox{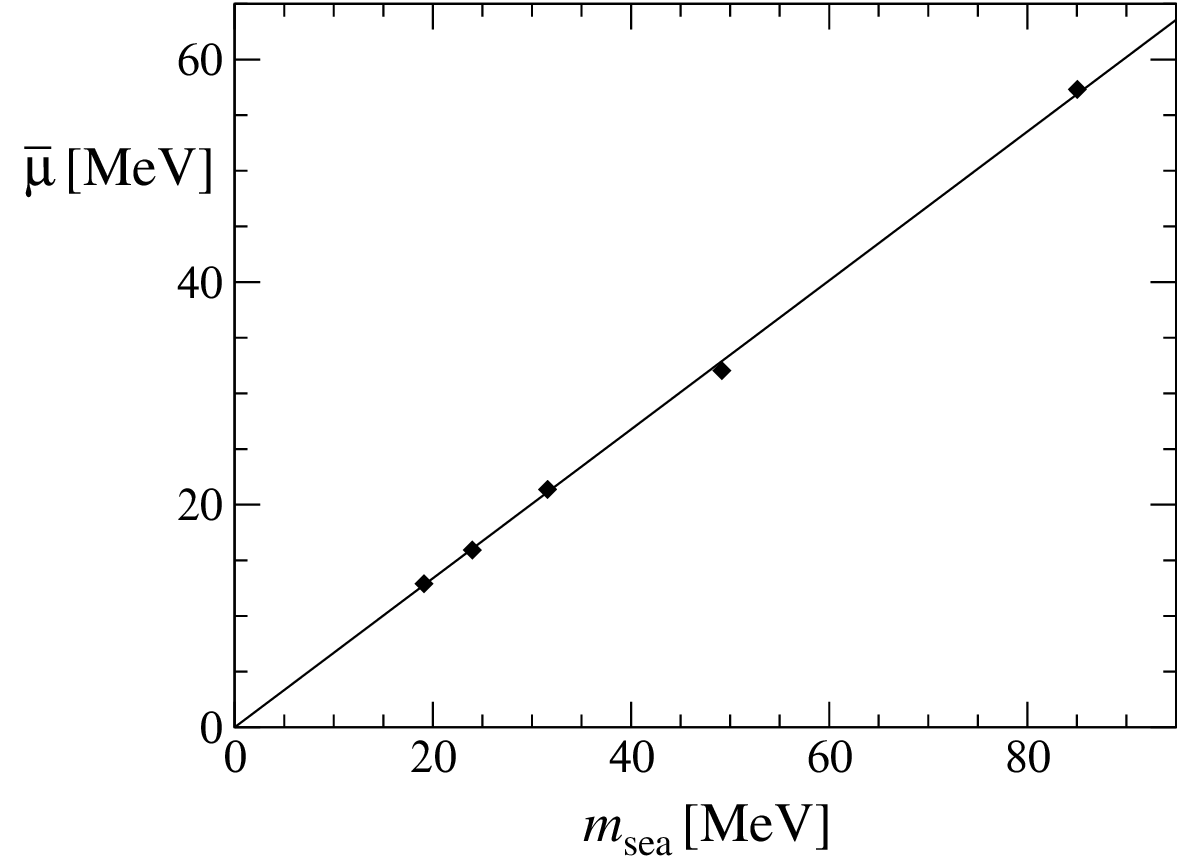}%
\vskip0.2cm
\figurecaption{%
Median $\mubar$ of the gap distributions
obtained in runs $D_1-D_5$ (data points), plotted as a function of 
the bare sea-quark mass $\msea$ 
(see subsect.~5.2 for the precise definition of the latter). The
line is a linear fit of the data without constant term.
}
\vskip0.0cm
}
\endinsert

However, the dependence of the width $\sigma$ of the distributions 
on the quark mass and the lattice size is different
(see table~3)%
$\kern1.5pt$\footnote{$\dagger$}{\footnotefont%
Following ref.~[\ref{Stability}], we define the width of the distributions
through $\sigma=\frac{1}{2}(v-u)$, where $[u,v]$ is the smallest range
in $\mu$, which contains more than $68.3\%$ of the data.}. 
In the case
of the $D$-series of lattices, for example, the width decreases by
as much as a factor of $1.5$ from the largest to the smallest quark
mass, while no obvious mass-dependence was seen
on the $A$ and $B$ lattices.
Moreover, $\sigma$ does not 
appear to scale proportionally to 
the inverse square root of the (four-dimensional) volume $V$ of the
lattice (see fig.~3). 
The widths on the lattices
$D_2$ and $E_2$, for example, turned out to be nearly the same,
contrary to what was expected on the basis of the experience
made in the unimproved theory.
  
Another perhaps not unrelated observation is that 
the median of the distribution on the $D$ and $E$ lattices
is always smaller than
the threshold of the spectral density in infinite volume,
which we expect to be at $\ZA\msea$ [\ref{Stability}], 
$\ZA$ being 
the axial-current renormalization constant
($\ZA=0.75(1)$ on these lattices [\ref{ZAalpha}]).
The spectral density in finite volume thus has a tail 
that extends a few MeV below the threshold. On the other
hand, the values quoted in table~3 
of the average splitting $\langle\Delta\rangle$ 
of the lowest four eigenvalues suggest that the 
tail scales to zero in the infinite-volume limit,
as it has to be if the density in infinite volume 
does not extend all the way to zero [\ref{Stability}].

\topinsert
\newdimen\digitwidth
\setbox0=\hbox{$0$}
\digitwidth=\wd0
\catcode`@=\active
\def@{\kern\digitwidth}
\tablecaption{Median and width of the gap distributions
in the improved theory}
\vskip1.0ex

$$\vbox{\settabs\+&%
                  xxxxxx&x&
                  xxxxxxxxxxx&xx&
                  xxxxxxxxxxx&xx&
                  xxxxxxxxxxx&xx&
                  xxxxxxxxxxx&xx\cr
\thicktablerule
\vskip1.0ex
                \+& \hfill Run\hfill
                 && \hfill $\mubar$\hfill
                 && \hfill $\sigma$\hfill
                 && \hfill $\mubar-\ZA\msea$\hfill
                 && \hfill $\langle\Delta\rangle$\hfill
                 &\cr
\vskip1.0ex
\thintablerule
\vskip1.5ex
  \+& \hskip2.5ex $D_1$\hfill
  &&  \hfill $57.3(6)$\hfill
  &&  \hfill $3.3(4)@@$\hfill 
  &&  \hfill $-6.5(10)$\hfill
  &&  \hfill $2.48(12)$\hfill
  &\cr
\vskip0.3ex
  \+& \hskip2.5ex $D_2$\hfill
  &&  \hfill $32.0(3)$\hfill
  &&  \hfill $2.79(24)$\hfill 
  &&  \hfill $-4.8(6)@$\hfill
  &&  \hfill $2.39(7)@$\hfill
  &\cr
\vskip0.3ex
  \+& \hskip2.5ex $D_3$\hfill
  &&  \hfill $21.4(3)$\hfill
  &&  \hfill $2.84(23)$\hfill 
  &&  \hfill $-2.3(4)@$\hfill
  &&  \hfill $2.29(7)@$\hfill
  &\cr
\vskip0.3ex
  \+& \hskip2.5ex $D_4$\hfill
  &&  \hfill $15.9(3)$\hfill
  &&  \hfill $2.33(18)$\hfill 
  &&  \hfill $-2.1(3)@$\hfill
  &&  \hfill $2.23(6)@$\hfill
  &\cr
\vskip0.3ex
  \+& \hskip2.5ex $D_5$\hfill
  &&  \hfill $12.9(4)$\hfill
  &&  \hfill $1.99(15)$\hfill 
  &&  \hfill $-1.4(4)@$\hfill
  &&  \hfill $2.28(5)@$\hfill
  &\cr
\vskip0.3ex
  \+& \hskip2.5ex $E_2$\hfill
  &&  \hfill $30.3(3)$\hfill
  &&  \hfill $2.58(19)$\hfill 
  &&  \hfill $-6.6(6)@$\hfill
  &&  \hfill $1.69(8)@$\hfill
  &\cr
\vskip0.3ex
  \+& \hskip2.5ex $E_3$\hfill
  &&  \hfill $21.3(3)$\hfill
  &&  \hfill $2.31(19)$\hfill 
  &&  \hfill $-5.8(5)@$\hfill
  &&  \hfill $1.52(7)@$\hfill
  &\cr
\vskip1.5ex
\thicktablerule
\vskip1.5ex
\+&{\footnotefont\kern2pt All entries are given in MeV}&\cr
}
$$
\vskip0.0ex
\endinsert

At present, however, there is still 
no theoretical understanding of the 
dependence of the gap distribution on the 
quark mass and the lattice size.
In particular, the fact that the improved and the unimproved
theory behave differently in this respect remains unexplained.
Partially quenched (Wilson) chiral perturbation theory 
may be a framework in which these questions can be addressed 
[\ref{SharpeGap}]
and further insight may perhaps also be gained by studying the localization
properties of the eigenfunctions and the convergence of the spectral density
to the infinite-volume limit.
It would be interesting to know, for example, whether the 
spectral gap coincides with the mobility edge [\ref{GoltermanMobility}]
and whether the 
tail of the spectral density below
$\ZA\msea$ does in fact disappear in the 
infinite-volume limit.

\subsection 3.2 Accessible range of pion masses on the $D$ and $E$ lattices

When the sea-quark mass decreases, the gap distribution 
becomes sharper and moves closer to the origin.
Eventually the probability for exceptionally small eigenvalues 
is not completely negligible anymore and one may run into algorithmic 
instabilities. We have not reached this point yet and 
consequently cannot say in which way
the DD-HMC simulations will be affected.
However, in order to be on the safe side, one may prefer to 
stay in the range of parameters where the gap distribution 
is well separated from the origin, i.e.~where, say, the 
inequality $\mubar\geq3\sigma$ holds [\ref{Stability}].

On a given lattice, this bound sets a lower limit on the accessible
sea-quark masses and thus on the masses $\Mpi$ of the 
pions (the lightest pseudo-scalar mesons made of the sea quarks). 
Furthermore, if large finite-volume effects are to be avoided, 
the bound $\Mpi L\geq3$ (where $L$ denotes the spatial lattice size)
should better be respected as well.

In the case of the $D$ and $E$ lattices, the range of pion
masses where both conditions are fulfilled can be determined
explicitly, using our simulation results. 
An extra\-polation in the sea-quark mass is however still required,
but it seems reasonable to extrapolate
$\mubar$ and $\Mpi^2$ linearly [\ref{I}]
and to assume that $\sigma$ drops to values below $2$~MeV 
at small quark masses.
For the accessible range of pion masses, we then obtain
\equation{
   \Mpi\geq\cases{\hbox{$314$ MeV} & ($D$ lattices),\cr
                  \noalign{\vskip1ex}
                  \hbox{$270$ MeV} & ($E$ lattices),\cr}
   \enum
}
where the limit is set by the constraint $\Mpi L\geq3$ 
on the $D$ lattices. This is not so on the $E$ lattices,
but values of $\Mpi L$ as low as $3.4$ can 
still be safely reached, i.e.~also in this case, the stability
bound is not too restrictive.

\topinsert
\vbox{
\vskip0.0cm
\epsfysize=4.4cm\hskip0.6cm\epsfbox{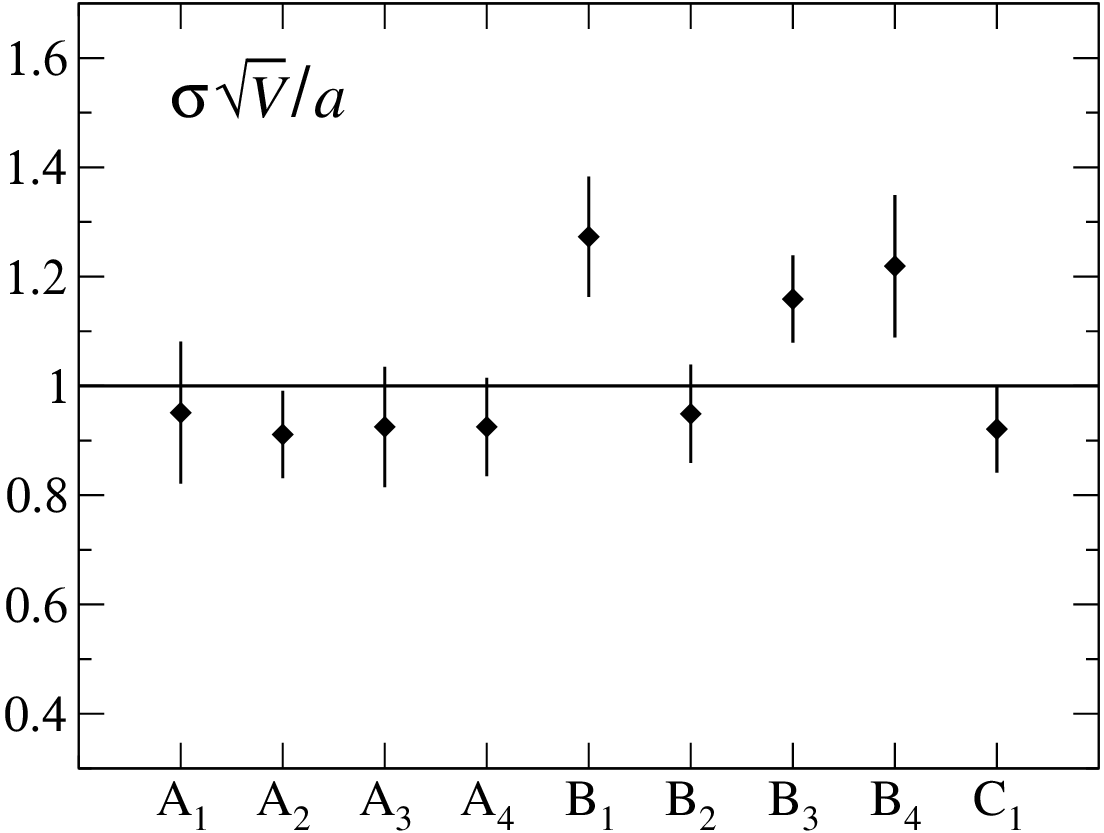}%
\epsfysize=4.4cm\hskip0.5cm\epsfbox{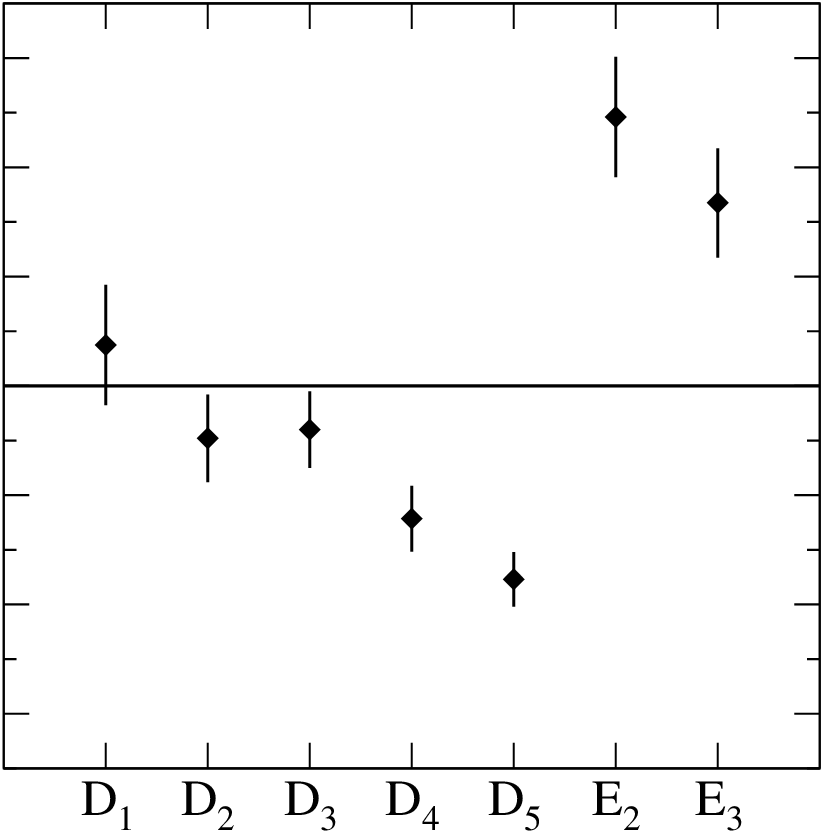}
\vskip0.3cm
\figurecaption{%
Width $\sigma$ of the gap distributions, scaled by the factor $\sqrt{V}/a$,
as obtained in the unimproved (left) and the improved theory (right).
The statistical errors were determined using the bootstrap method.
}
\vskip0.0cm
}
\endinsert

\subsection 3.3 Molecular-dynamics instabilities

Similar to the standard HMC algorithm,
the DD-HMC algorithm
obtains the next field configuration by integrating 
the associated molecular-dynamics equations.
The numerical integration of these
equations is well known to be potentially unstable.
If an instability occurs,
the energy deficit $\Delta H$ at
the end of the integration can be large  
and the new field configuration is then normally rejected.
The efficiency of the simulation may thus be affected, 
but we wish to emphasize that large energy deficits do 
not invalidate the algorithm unless the reversibility
of the molecular-dynamics integration is compromised.

Earlier studies of the phenomenon suggest
that the instabilities are caused by
exceptionally small eigenvalues of the lattice Dirac operator
[\ref{UKQCDlight}--\ref{ALPHAstability}].
Even if the gap distribution 
is safely separated from zero, it is possible
that the Dirac operator develops such eigenvalues 
somewhere along the molecular-dynamics 
trajectories. The probability for this
depends on how accurately the molecular-dynamics equations 
are solved, i.e.~on the integration step sizes and 
the solver residues.

In our simulations, the probability for 
$|\Delta H|$ to be larger than $2$ was always fairly small and often 
equal to zero (runs $B_1-B_4$, for example). The worst
cases in the unimproved and the improved theory
were the runs $A_4$ and $D_5$ respectively,
where the threshold
of $2$ was passed by $1.4\%$ and 
$0.7\%$ of the trajectories.
Energy deficits $|\Delta H|$
larger than $10^3$ were never seen, but values above
$100$ did occur, although very rarely so.

\section 4. Autocorrelation times

The dynamical properties 
of the simulation algorithms used in lattice
QCD are still largely unknown. It is not clear, for example,
whether there are several relevant time scales and how
they depend on the lattice parameters and the chosen algorithm.
We shall not attempt to answer these difficult questions here, however,
and merely give an account of our empirical studies of 
the autocorrelations in the runs listed in table~1.

\subsection 4.1 Determination of autocorrelation times

Following the standard conventions, we define
the integrated autocorrelation time $\tauint$ of an infinite series 
$a_1,a_2,a_3,\ldots$ of measured values of an observable $A$ through
\equation{
  \tauint={1\over2}+\sum_{t=1}^{\infty}{\Gamma(t)\over\Gamma(0)},
  \enum
}
where $\Gamma(t)$ denotes the autocorrelation function of the series.
In practice only a finite number $N$ of measurements can be made and 
the estimation of the autocorrelation time from the available data
then requires some ad hoc choices to be made. 

For the autocorrelation function we use the approximation
\equation{
  \Gamma(t)\simeq
  {1\over N-t}\sum_{i=1}^{N-t}(a_i-\abar_{-})(a_{i+t}-\abar_{+}),
  \qquad 0\leq t<N,
  \enum
}
in which $\abar_{-}$ and $\abar_{+}$ are, respectively, the averages 
of the first $N-t$ and the last $N-t$ 
elements of the series $a_1,\ldots,a_N$.
The sum in eq.~(4.1) is then
truncated at some value $W\ll N$ of the time lag $t$, referred to 
as the summation window, which should ideally be such that 
the remainder of the sum can be safely neglected.

\topinsert
\vbox{
\vskip0.0cm
\epsfysize=7.0cm\hskip2.0cm\epsfbox{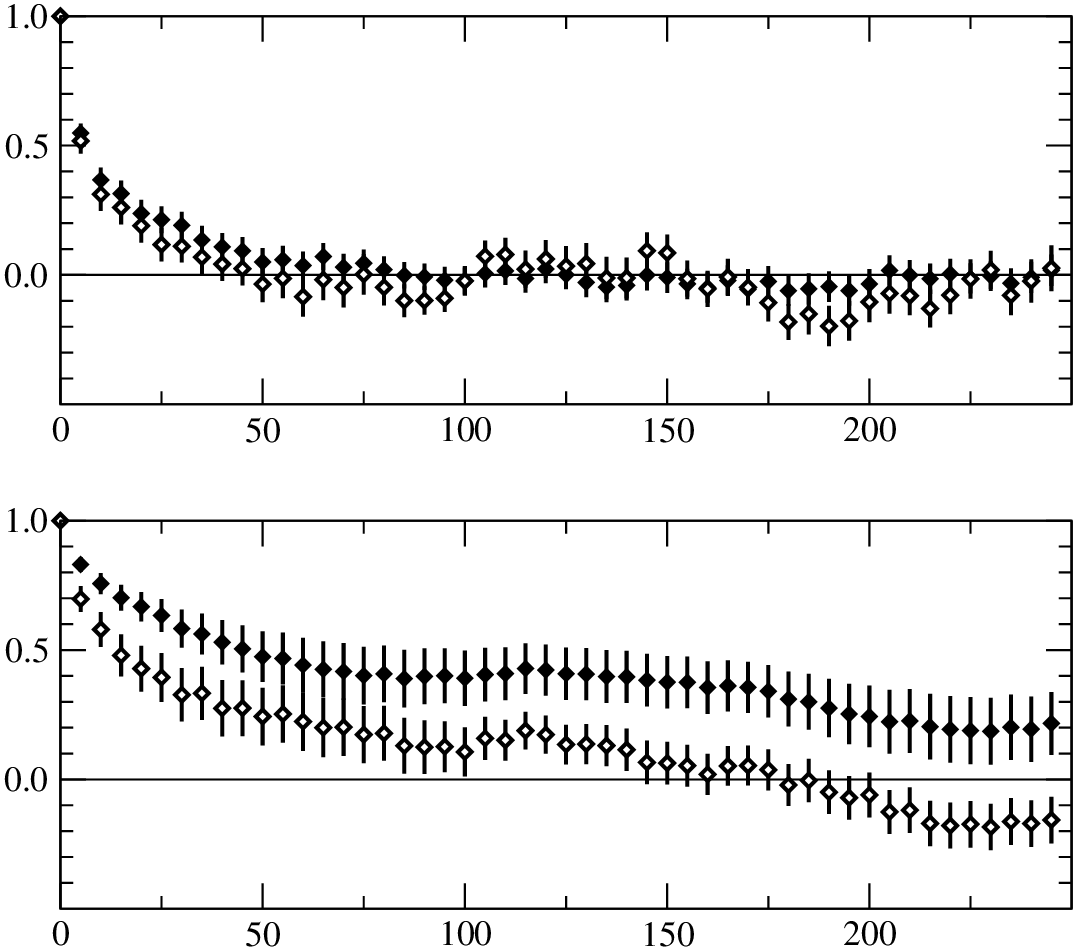}%
\vskip0.3cm
\figurecaption{%
Normalized autocorrelation functions $\Gamma(t)/\Gamma(0)$,
plotted versus the time lag $t$ given in numbers of trajectories,
of the plaquette $P$ (upper plot) and the solver iteration 
number $\Ngcr$ (lower plot).
The data shown were calculated using
the last $4000$ trajectories of run $B_2$ (full points) or only the 
first $2000$ of these (open points).
}
\vskip0.0cm
}
\endinsert

If the autocorrelation function is well behaved,
as in the case shown in the upper plot of fig.~4,
the choice of the summation window is not 
critical and any reasonable prescription will do.
The rule adopted here is to stop the summation in eq.~(4.1)
at the first value of $t$ where the normalized autocorrelation function
is equal to zero within two times its statistical error, the latter
being estimated using the Madras-Sokal approximation 
(see appendix E of ref.~[\ref{SchwarzIII}]).

In practice the calculated autocorrelation functions may 
have long tails and they may also vary significantly 
with the selected range of the data series. 
An example illustrating 
this behaviour is shown in the lower plot in fig.~4.
In all these cases, we divide the data series into
large bins, calculate the bin averages
and estimate the statistical variance $\sigma^2$ of the 
total average assuming these are statistically independent.
The integrated autocorrelation time is then given by
\equation{
   \tauint={\sigma^2\over2\sigma_0^2},
   \enum
}
where $\sigma_0$ denotes the naive statistical error.
Evidently, the results obtained in this way 
are rough estimates that could easily be wrong by factor $2$ or so.

\subsection 4.2 Reference autocorrelation times

The integrated autocorrelation times
of the Wilson plaquette $P$ and the
GCR solver iteration number $\Ngcr$ are listed
in table~4. These two quantities are unphysical, 
but they are readily accessible and are useful reference cases 
that probe the
dynamics of the simulation at both short and long distances.

\topinsert
\newdimen\digitwidth
\setbox0=\hbox{$0$}
\digitwidth=\wd0
\catcode`@=\active
\def@{\kern\digitwidth}
\tablecaption{Autocorrelation times of the plaquette $P$
and the solver iteration number $\Ngcr$} 
\vskip-1.0ex

$$\vbox{\settabs\+&%
                  xxxxxx&x&
                  xxxxxxxxxxx&x&
                  xxxxxxxxxxx&xxxxxxx&
                  xxxxxx&x&
                  xxxxxxxxxxx&x&
                  xxxxxxxxxxx&xx\cr
\thicktablerule
\vskip1.0ex
                \+& \hfill Run\hfill
                 && \hfill $\tauint$[{\ninepoint$P$}]\hfill
                 && \hfill $\tauint$[{\ninepoint$\Ngcr$}]\hfill
                 && \hfill Run\hfill
                 && \hfill $\tauint$[{\ninepoint$P$}]\hfill
                 && \hfill $\tauint$[{\ninepoint$\Ngcr$}]\hfill
                 &\cr
\vskip1.0ex
\thintablerule
\vskip1.5ex
  \+& \hskip2.5ex $A_{1a}$\hfill
  &&  \hfill $25(5)$\hfill
  &&  \hfill $43^*\phantom{(0}$\hfill 
  &&  \hskip2.5ex $C_{1}$\hfill
  &&  \hfill $17(3)$\hfill
  &&  \hfill $35(7)$\hfill 
  &\cr
\vskip0.3ex
  \+& \hskip2.5ex $A_{1b}$\hfill
  &&  \hfill $29(6)$\hfill
  &&  \hfill $38^*\phantom{(0}$\hfill 
  &&  \hskip2.5ex $D_{1}$\hfill
  &&  \hfill $11(1)$\hfill
  &&  \hfill $10(2)$\hfill 
  &\cr
\vskip0.3ex
  \+& \hskip2.5ex $A_{2}$\hfill
  &&  \hfill $23(4)$\hfill
  &&  \hfill $46^*\phantom{(0}$\hfill 
  &&  \hskip2.5ex $D_{2}$\hfill
  &&  \hfill $17(3)$\hfill
  &&  \hfill $21(4)$\hfill 
  &\cr
\vskip0.3ex
  \+& \hskip2.5ex $A_{3a}$\hfill
  &&  \hfill $14(2)$\hfill
  &&  \hfill $\phantom{0}53(10)$\hfill 
  &&  \hskip2.5ex $D_3$\hfill
  &&  \hfill $16(2)$\hfill
  &&  \hfill $19(3)$\hfill 
  &\cr
\vskip0.3ex
  \+& \hskip2.5ex $A_{3b}$\hfill
  &&  \hfill $28^*\phantom{()}$\hfill
  &&  \hfill $53^*\phantom{(0}$\hfill 
  &&  \hskip2.5ex $D_4$\hfill
  &&  \hfill $16(2)$\hfill
  &&  \hfill $15(2)$\hfill 
  &\cr
\vskip0.3ex
  \+& \hskip2.5ex $A_{4}$\hfill
  &&  \hfill $19(4)$\hfill
  &&  \hfill $45^*\phantom{(0}$\hfill 
  &&  \hskip2.5ex $D_{5}$\hfill
  &&  \hfill $32(6)$\hfill
  &&  \hfill $24(5)$\hfill 
  &\cr
\vskip0.3ex
  \+& \hskip2.5ex $B_{1}$\hfill
  &&  \hfill $14(2)$\hfill
  &&  \hfill $50^*\phantom{(0}$\hfill 
  &&  \hskip2.5ex $E_{1}$\hfill
  &&  \hfill $33^*\phantom{()}$\hfill
  &&  \hfill $14(3)$\hfill 
  &\cr
\vskip0.3ex
  \+& \hskip2.5ex $B_{2}$\hfill
  &&  \hfill $12(2)$\hfill
  &&  \hfill $39^*\phantom{(0}$\hfill 
  &&  \hskip2.5ex $E_{2}$\hfill
  &&  \hfill $19(3)$\hfill
  &&  \hfill $11(2)$\hfill 
  &\cr
\vskip0.3ex
  \+& \hskip2.5ex $B_{3}$\hfill
  &&  \hfill $@9(1)$\hfill
  &&  \hfill $45^*\phantom{(0}$\hfill 
  &&  \hskip2.5ex $E_{3}$\hfill
  &&  \hfill $27(5)$\hfill
  &&  \hfill $25(5)$\hfill 
  &\cr
\vskip0.3ex
  \+& \hskip2.5ex $B_{4}$\hfill
  &&  \hfill $14(2)$\hfill
  &&  \hfill $51^*\phantom{(0}$\hfill 
  &&  \hskip2.5ex$$\hfill
  &&  \hfill $$\hfill
  &&  \hfill $$\hfill 
  &\cr
\vskip1.5ex
\thicktablerule
\vskip1.5ex
\+&{\footnotefont$^*$ Estimate based on data binning}&\cr
}
$$
\vskip-2ex
\endinsert

In order to facilitate the comparison of the figures quoted in the 
table, the autocorrelation times were determined using data
series of a fixed length equal to $4000$ trajectories. 
The autocorrelation times are given
in numbers of trajectories and error estimates are
quoted only in those cases where the autocorrelation function was 
well behaved. 
In these regular situations, the binning method
always gave consistent results.

In all simulations of the improved theory, except for run $E_1$ perhaps,
the autocorrelation times
were safely determined and turned out to be reasonably small.
This was not so in the simulations of the 
unimproved theory, where the autocorrelation
function of the GCR iteration number typically had a tail
similar to the one shown in the lower plot in fig.~4.
$\rmO(a)$ improvement thus appears to have
the side-effect of reducing the autocorrelation times.

The regularity of run $C_1$ then remains unexplained, however,
and the differences in the autocorrelation times could actually
also very well be related to the fact
that the physical volumes of the $C_1-E_3$ lattices are
larger, by a factor of two or more, than the volumes of the other lattices.
Presumably the size of the blocks, on which the DD-HMC algorithm operates,
matters as well, although the comparison of the runs $A_{1a}$ and 
$A_{1b}$ does not suggest this to be so.

\subsection 4.3 Autocorrelations of physical quantities

The meson masses and all other physical quantities
were calculated after finishing the simulations, using the generated
ensembles of saved gauge-field configurations
(see table~1). A fairly large number of trajectories
was skipped between successive saved configurations so that the statistical 
correlations in these sets of fields can be expected to be small.

In order to find out whether the residual correlations 
are relevant for the determination of the statistical errors, 
the basic two-point correlation functions
were averaged over small bins of successive configurations.
The physical quantities were then extracted from the binned data
and their statistical errors were estimated using 
the jackknife method (appendix A).
If there were significant statistical correlations
in the data,
the errors would increase with the bin size, but 
this was not the case and we therefore concluded that 
it was safe to proceed without data binning.

\section 5. Computation of meson masses and decay constants

The masses and matrix elements tabulated in appendix C
were calculated using a combination of methods,
most of which being entirely standard by now. 
We consider two valence quarks, labelled $r$ and $s$,
and study the vector and pseudo-scalar mesons in the $\rbar s$-channel. 
The masses of the valence quarks 
may be set to the sea-quark mass, but we are
also interested in the partially quenched situation where one
of the quark masses is different from the sea-quark mass.

\subsection 5.1 Two-point correlation functions

The pseudo-scalar density, the axial current and the vector current
in the $\rbar s$-channel are given by
\equation{
   P^{rs}=\rbar\dirac{5}s,
   \qquad
   A_{\mu}^{rs}=\rbar\dirac{\mu}\dirac{5}s,
   \qquad
   V_{\mu}^{rs}=\rbar\dirac{\mu}s.
   \enum
}
All masses and decay constants we are interested in 
were extracted from the two-point functions
\equation{
  \fpp(x_0)=a^3\sum_{x_1,x_2,x_3}
  \left\langle P^{rs}(x)P^{sr}(0)\right\rangle,
  \enum
  \next{2ex}
  \fap(x_0)=a^3\sum_{x_1,x_2,x_3}
  \left\langle A_0^{rs}(x)P^{sr}(0)\right\rangle,
  \enum 
  \next{2ex}
  \fvv(x_0)=a^3\sum_{x_1,x_2,x_3}\sum_{k=1}^3
  \left\langle W_k^{rs}(x)W_k^{sr}(0)\right\rangle,
  \enum
}
where $W_{\mu}^{rs}$ is 
a linear combination of the vector current $V^{rs}_{\mu}$ and 
a Jacobi smeared form of it [\ref{Jacobi}], 
slightly tuned so as to suppress the high-energy intermediate
states
in the two-point function.

The correlation functions were evaluated in the standard manner
by first expressing them as an expectation value of a 
product of two quark propagators.
These were calculated 
by solving the lattice Dirac equation,
using the Schwarz-preconditioned GCR solver [\ref{SchwarzII}]
and requiring the 
normalized residue of the solution to be less than $10^{-10}$.
In order to reduce the statistical fluctuations,
the results were averaged over time-reflections and $5$ 
distant source points in the case of the 
$A$ and $B$ runs and over $3$ source points in the case 
of the $D$ runs.

\subsection 5.2 Masses and matrix elements

On a lattice of infinite time-like extent,
and at large times $x_0$,
the correlation function $\fpp(x_0)$ is saturated by
the one-particle pseudo-scalar meson state in the $\rbar s$-channel. 
If we denote the mass of the meson by $\Mps$ and the associated
vacuum-to-meson matrix element by $\Gps$,
the asymptotic form of the correlation function is
\equation{
  \fpp(x_0)=-{\Gps^2\over\Mps}\rme^{-\Mps x_0}+\ldots,
  \enum
}
where the ellipsis stands for a series of more rapidly decaying terms.
The mass $\Mv$ of the $\rbar s$ vector meson may be defined similarly
through the asymptotic behaviour of the vector correlation function
$\fvv(x_0)$,
but the definition requires further explanation if the meson is unstable
in infinite volume (see subsect.~5.6).

Next we note that the ratio
\equation{
   \meff(x_0)=\left\{\frac{1}{2}\left(\drv{0}+\drvstar{0}\right)\fap(x_0)+
   \cA a\drvstar{0}\drv{0}\fpp(x_0)\right\}\bigm/\fpp(x_0)
   \enum
}
converges to a constant $m_{rs}$ at large times $x_0$,
for any fixed value of the parameter $\cA$, 
because both $\fap(x_0)$ and $\fpp(x_0)$ are proportional
to $\rme^{-\Mps x_0}$ in this limit.
Moreover, in the continuum limit,
$\meff(x_0)$ is expected to converge to the sum of
the bare current-quark masses of the $r$ and the $s$ quark,
at all times $x_0$,
with a rate proportional to $a$
in the unimproved theory (where we set $\cA$ to zero) 
or $a^2$ if the improvement 
coefficients $\csw$ and $\cA$ are
properly tuned [\ref{OaImp},\ref{NPimp},\ref{NPimpCurrent}]%
\kern1.5pt\footnote{$\dagger$}{\footnotefont%
The effects of the $1+\rmO(am)$ renormalization
factors (C.2) are expected to be small in practice and are neglected
here for simplicity.
}.

All our numerical data for $\meff(x_0)$ in fact
turned out to be statistically consistent with a constant value, 
over a large range of $x_0$, and the quark mass sum $m_{rs}$ 
was therefore always unambiguously and accurately determined.
In particular, the current-quark mass $\msea=\frac{1}{2}m_{rr}$ 
of the sea quarks is obtained by 
setting the hopping parameters of the valence quarks to $\ksea$.
Whether in general $m_{rs}$ coincides with 
$\frac{1}{2}(m_{rr}+m_{ss})$, as one expects to be the case if
the lattice effects are small, is a question 
to which we shall return in sect.~6. 

The bare pseudo-scalar decay constant $\Fps$ in the $\rbar s$-channel
is normally extracted from the asymptotic behaviour of 
the two-point functions $\fap(x_0)$ and $\fpp(x_0)$.
In this paper, however, we first computed $m_{rs}$, $\Mps$ and $\Gps$ 
and then used the formula
\equation{
  \Fps={m_{rs}\over\Mps^2}\Gps
  \enum
}
for the decay constant. Starting
from eq.~(5.6), it is straightforward to show that
equivalent results are obtained in this way, up to
small corrections of O($a^2$). 
Note that $\Fps$ is automatically O($a$)-improved
if $m_{rs}$ is.

\subsection 5.3 Spectral decomposition in finite volume

On a finite lattice with time-like extent $T$, the 
calculation of the pseudo-scalar and vector meson masses
requires some care and must address the issue of
higher-states contributions. This is, incidentally, not
so in the case of the quark mass sum $m_{rs}$, which 
is expected to be independent of the lattice size up
to lattice-spacing effects.

For $0<x_0<T$,
the correlation function $\fpp(x_0)$ (and similarly 
$\fvv(x_0)$) can be expanded in a
rapidly convergent series of the form
\equation{
  \fpp(x_0)=-\sum_{i=0}^{\infty}\sum_{j=i}^{\infty}
  c_{ij}h(x_0;E_i,E_j),
  \enum
  \next{2ex}
  h(t;E,E')=\exp\{-Et-E'(T-t)\}+\exp\{-E't-E(T-t)\},
  \enum
}
where $0=E_0<E_1<E_2<\ldots$ are the intermediate-state energies
and $c_{ij}\geq0$ the associated spectral weights%
\kern1.5pt\footnote{$\ddagger$}{\footnotefont%
Equation (5.8) assumes the existence of a positive hermitian transfer
matrix which may not be guaranteed in the improved theory. 
It seems likely to us, however, that a transfer matrix can still
be defined,
as is the case in O($a^2$)-improved gauge theories
[\ref{LuscherWeiszTrans}], although complex energy values and negative
weights may occur at energies on the order of the cutoff scale $1/a$.
}.
In the channel considered here, the lowest intermediate state is
the $\rbar s$ pseudo-scalar meson state at zero spatial momentum. Then come
the multi-meson scattering states and more and more complicated 
states as one moves up the energy scale.

At large $x_0$ and $T$, the dominant term in the series (5.8) is
thus the one where $E_i=0$ and $E_j=\Mps$.
Moreover, using the product inequality (B.3), 
the contributions of all higher-energy states can be shown to be
exponentially suppressed with respect to this term.
In practice their effects are seen in the simulation data 
only when either $x_0$ or $T-x_0$ is not too large.
The leading terms in this range are then
\equation{
  \fpp(x_0)=c_0h(x_0;0,M_0)+c_1h(x_0;0,M_1)+\ldots,
  \qquad M_0=\Mps,
  \enum
}
where $M_1$ denotes the energy 
of the next-to-lowest state in the $\rbar s$-channel
(if the spatial volume of the lattice is large enough, this 
will be a three-meson state with all particles at rest).

Note that each term in the spectral series (5.8) 
decreases exponentially in the range
$0\leq x_0\ll\frac{1}{2}T$, with an exponent equal to $E_j-E_i$ that 
can be as small as the pseudo-scalar meson mass,
for example, even if both 
$E_i$ and $E_j$ are not small.
The presence of such contributions complicates the analysis
of the correlation functions considerably unless 
the time-like extent $T$ of the lattice is 
sufficiently large to strongly suppress them.
This condition was barely satisfied in the case of the run $A_4$,
which is why we decided to discard it from
the physics analysis (as already mentioned in the first paper in this 
series).

\subsection 5.4 Effective masses and matrix elements

Slightly departing from what is usually done,
we define the effective pseudo-scalar meson mass
$\Meff(x_0)$ in the $\rbar s$-channel to be the value of $M\geq0$ where
\equation{
  {h(x_0-a;0,M)\over h(x_0;0,M)}=
  {\fpp(x_0-a)\over\fpp(x_0)}.
  \enum
}
Using the results obtained in appendix B, it is not difficult to prove
that this equation has one and only one solution.
Moreover, with this definition of the effective mass it is 
guaranteed that $\Meff(x_0)=\Mps$ at large $x_0$, 
up to exponentially small terms.
We then also introduce the effective matrix element
\equation{
  \Geff(x_0)=
  \left\{-\Meff(x_0){\fpp(x_0)\over h(x_0;0,\Meff(x_0))}\right\}^{1/2},
  \enum
}
which converges to $\Gps$ in the large-time limit.

The asymptotic behaviour 
of the effective mass at large $x_0$ and $T$
can be worked out explicitly, starting from the 
spectral representation (5.10). 
Setting
\equation{
  \eps(x_0)={c_1h(x_0;0,M_1)\over c_0h(x_0;0,M_0)},
  \qquad
  \del(x_0)=\left\{M{\partial\over\partial M}\ln h(x_0;0,M)\right\}_{M=M_0},
  \enum
}
and going through a few lines of algebra,
it is straightforward to derive the expansion
\equation{
  \Meff(x_0)=\Mps\left\{1+{\eps(x_0)-\eps(x_0-a)\over\del(x_0)-\del(x_0-a)}
  +\ldots\right\},
  \enum
}
where the ellipsis stands for terms that are 
exponentially small with respect to the next-to-leading term.
A similar formula,
\equation{
  \Geff(x_0)=\Gps\left\{1+\frac{1}{2}\eps(x_0)+
  \frac{1}{2}\left(1-\delta(x_0)\right)
  {\eps(x_0)-\eps(x_0-a)\over\del(x_0)-\del(x_0-a)}+\ldots\right\},
  \enum
}
is obtained in the case of the effective matrix element.

\subsection 5.5 Fit procedures

From the point of view of the statistical error analysis,
the correlation functions $\fpp$, $\fap$ and $\fvv$
are the primary quantities, while the effective quark mass sums, 
meson masses and matrix elements are 
functions of these. 
The statistical errors of all these quantities
tend to be strongly correlated.
We took the correlations fully into account, 
from the primary quantities to the final results,
by propagating the errors using the jackknife method
(appendix A).
In particular, fitted and interpolated values were always
considered to be functions of the input data, 
which allows their errors to be calculated in the 
standard manner.

The quark mass sum, the pseudo-scalar 
meson masses and matrix elements, and the masses of the vector mesons were
all obtained by fitting the corresponding effective quantity $\Peff(x_0)$
in a range $t_0\leq x_0\leq t_1$ of time with the chosen
fit function $\Phi(x_0)$. 
We performed correlated least-squares fits, where
the values of the fit parameters were determined by minimizing 
\equation{
  \chi^2=
  \sum_{x_0,y_0=t_0}^{t_1}
  \left[\Peff(x_0)-\Phi(x_0)\right]
  (C^{-1})_{x_0y_0}
  \left[\Peff(y_0)-\Phi(y_0)\right],
  \enum
}
the matrix $C$ being the statistical error covariance of 
$\Peff(t_0),\ldots,\Peff(t_1)$.
The quark mass sum $m_{rs}$, for example, was computed by
fitting $\meff(x_0)$ to a constant as shown in fig.~5a.

\topinsert
\vbox{
\vskip0.0cm
\epsfysize=8.0cm\hskip0.1cm\epsfbox{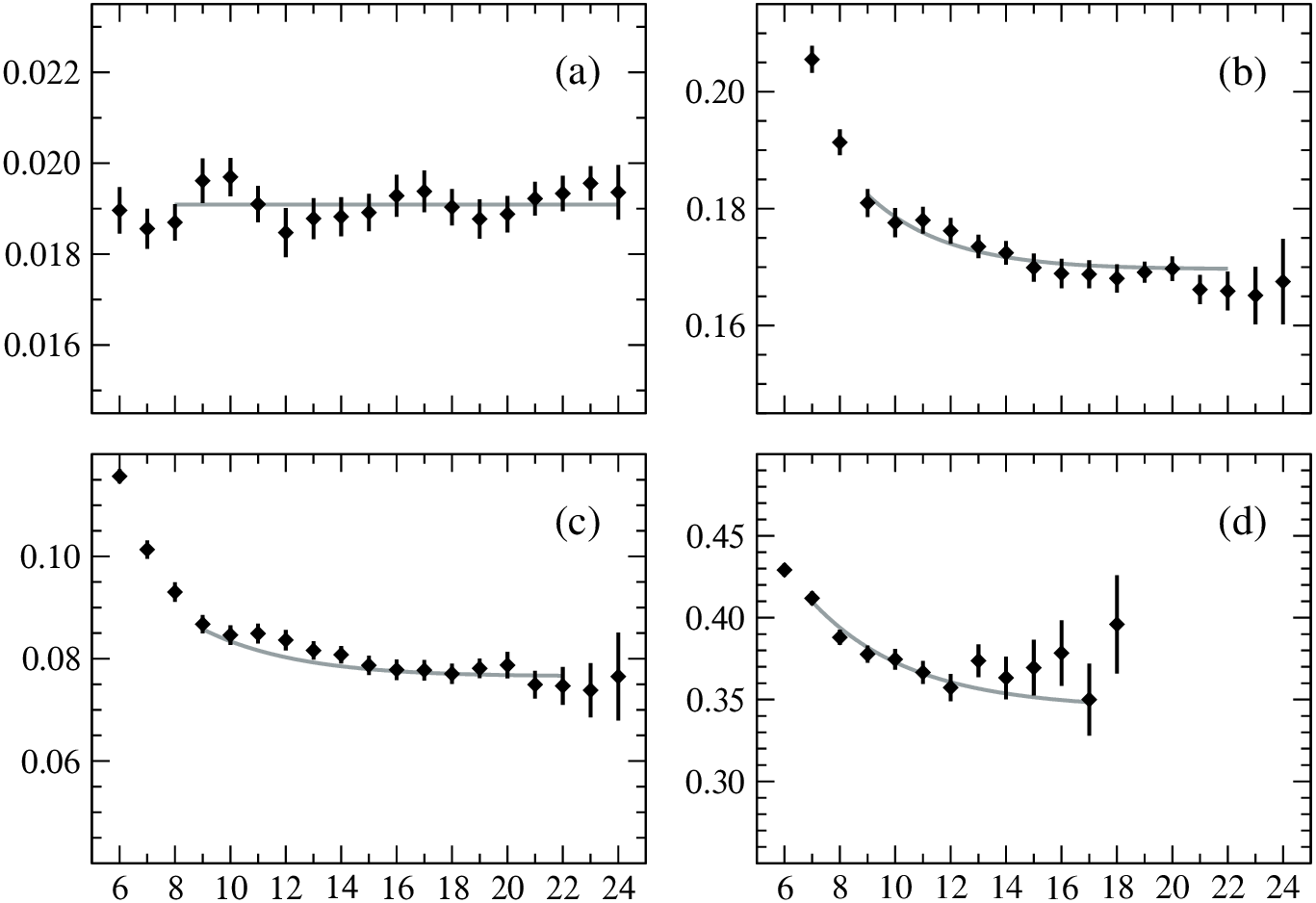}%
\vskip0.3cm
\figurecaption{%
Sample plots illustrating the dependence on $x_0/a$ of 
the effective quark mass sum (figure a),
the pseudo-scalar mass and matrix element (figures b and c) and
the vector meson mass (figure d), all given in lattice units.
The data points shown are from run $D_4$ and the valence quark masses 
were both set to the sea-quark mass in this example.
The solid lines are the fits discussed in the text.
}
\vskip0.0cm
}
\endinsert

In the case of the pseudo-scalar meson masses, we
fitted the data with the asymptotic
expression (5.14).
We first calculated the mass $\Mpi$ of the pions, i.e.~the mesons made of the 
sea-quarks, by substituting $M_1=3\Mpi$ for the energy of the next-higher 
state (thus assuming the latter is a three-pion
state with small interaction energy) and adjusting 
$\Mpi$ and $c_1/c_0$ so as to minimize $\chi^2$. 
While the fit curves obtained in this way represent the data
very well, it should be noted that the fitted value of $\Mpi$ is
largely determined by the data at large times $x_0$, where 
a fit to a constant would give nearly the same results
(see fig.~5b).

Once $\Mpi$ was determined, the mesons made of a sea quark and 
a valence quark with a mass different from the sea quark 
were considered. Here we set 
$M_1=\Mps+2\Mpi$ and otherwise proceeded as in the degenerate case.
Next the matrix elements $\Gps$ were computed by fitting the data
with the asymptotic expression (5.15), using the same
values of $M_1$ as in the fits of the effective meson 
masses (fig.~5c). We did not set $c_1/c_0$ to the previously 
computed values, but it turned out that 
the two fits gave consistent results for this parameter.

\subsection 5.6 Energy spectrum in the vector channel

At small sea-quark masses, the vector mesons become
resonances that decay into two (or more) pseudo-scalar mesons.
As was shown long ago [\ref{LuscherResonances}],
resonances give rise to a characteristic volume-dependent 
pattern of the energy spectrum 
which allows their masses and decay widths to be determined, in principle, 
from simulation data.

As before, we considered the channels 
where one or both of the $r$ and $s$ quarks is a sea quark.
Starting from the correlation functions $\fvv(x_0)$,
the lowest energy $\Mv$ in this channel was calculated
by fitting the effective mass with the asymptotic formula (5.14)
(with $\Mps$ replaced by $\Mv$). For the lowest excited-state energy
we substituted
\equation{
  M_1=(\Mps^2+k^2)^{1/2}+(\Mpi^2+k^2)^{1/2}, 
  \qquad k=2\pi/L,
  \enum
}
in this case, $L$ being the spatial size of the lattice.
Excellent fits were obtained with this ansatz 
and $\Mv$ was determined quite accurately on all lattices.

We refer to the energy values $\Mv$
as the vector meson masses in this paper,
even in those cases where the meson is likely to 
become a resonance in the infinite volume limit 
(we estimate this to be so at the lightest quark masses in each
series of lattices and perhaps
at some of the second-to-lightest as well).
This use of language is only slightly incorrect, however,
because in all our simulations 
$\Mv$ turned out to be at most $20\%$ larger than
$\Mps+\Mpi$ and significantly smaller than $M_1$,
in which case the true resonance energy is 
expected to be close to $\Mv$ [\ref{LuscherResonances}].

We finally note that the statistical errors in the vector channel
tend to be larger than those in the pseudo-scalar channel.
The effect could be related to the resonance character
of the vector mesons and it is conceivable that
a coupled channel analysis, such as the one recently presented by
Aoki et al.~[\ref{AokiResonances}], 
will not only allow the vector meson decays to be studied 
but may also help to reduce the statistical errors.

\section 6. Quark-mass dependence in partially quenched QCD

The most important physical results of our simulations were already
presented in our first paper in this series [\ref{I}]. We now 
discuss the dependence of the quantities tabulated in appendix C 
on the quark masses in some further detail, 
focusing on the empirical facts rather than
on their possible theoretical interpretation. 

As before we set $\msea=\frac{1}{2}m_{rr}$ if the $r$ quark is a sea quark and
we now also set $\mval=\frac{1}{2}m_{ss}$ if the 
$s$ quark is a valence quark. The figures in the tables are all for the mixed
case, where one quark is a sea quark and the other a valence quark.
We are thus considering partially 
quenched QCD with $2+1$ flavours of quarks.

\subsection 6.1 Quark and pseudo-scalar meson masses

We first remark that the quark mass sum $m_{rs}$
turns out to be equal to 
$\msea+\mval$ within statistical errors, on all lattices and for all
quark-mass combinations.
The ratio $m_{rs}/(\msea+\mval)$ is 
obtained with better statistical precision
than the quark masses, but the largest deviation
seen in this case is only $0.6\%$.
The additivity of the current quark masses
(which is an exact property of the theory in the continuum limit)
is thus accurately guaranteed on the lattices that we have simulated.

\topinsert
\vbox{
\vskip0.0cm
\epsfysize=9.0cm\hskip1.6cm\epsfbox{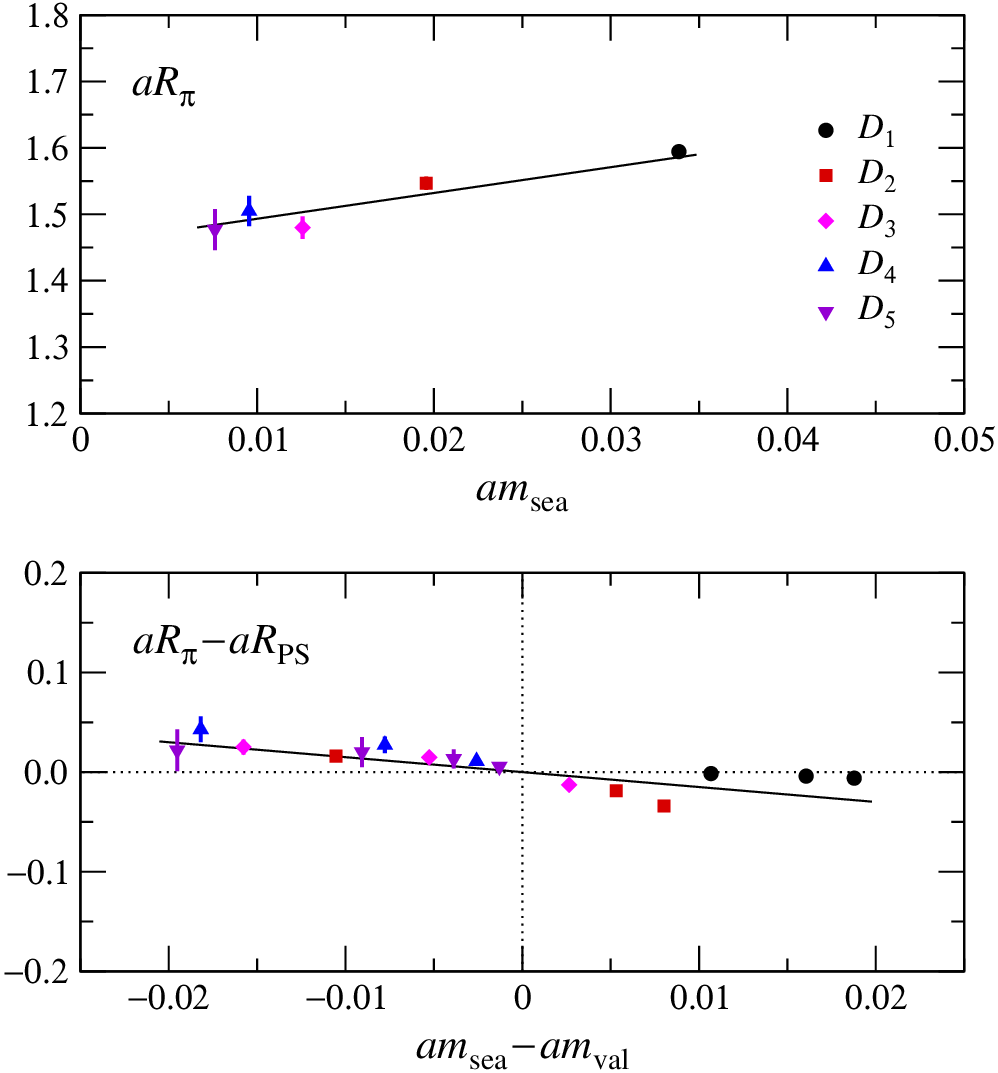}%
\vskip0.3cm
\figurecaption{%
Results for the ratio $\Rpi$ and the difference $\Rpi-\Rps$ 
obtained on the $D$-series of lattices. 
The solid lines represent the global linear fit (6.2).
Note that the points in the lower plot do not have to
line up within errors, since
$\Rpi-\Rps$ is a function of two independent variables
rather than of $\msea-\mval$ alone.
}
\vskip0.0cm
}
\endinsert

Next we consider the ratios
\equation{
  \Rps={\Mps^2\over \msea+\mval},
  \qquad
  \Rpi=\left.\Rps\right|_{\mval=\msea}={\Mpi^2\over2\msea},
  \enum
}
which are independent of the quark masses 
to lowest order of chiral perturbation theory. 
However, this is not so at next-to-leading order and 
the numerically calculated ratios are in fact
weakly mass-dependent (see fig.~6).
An empirical fit
\equation{
  \Rps=a_0+a_1(\msea+\mval)+a_2\msea
  \enum
}
represents the data quite well in the given range of masses
except perhaps for the points where $\mval\ll\msea$.
In the case of the $D$-series of lattices,
for example, the data for $\Rps$ deviate from the 
fit by no more than $2\%$ and 
most points are within a margin of $1\%$.

\topinsert
\vbox{
\vskip0.0cm
\epsfysize=9.0cm\hskip1.5cm\epsfbox{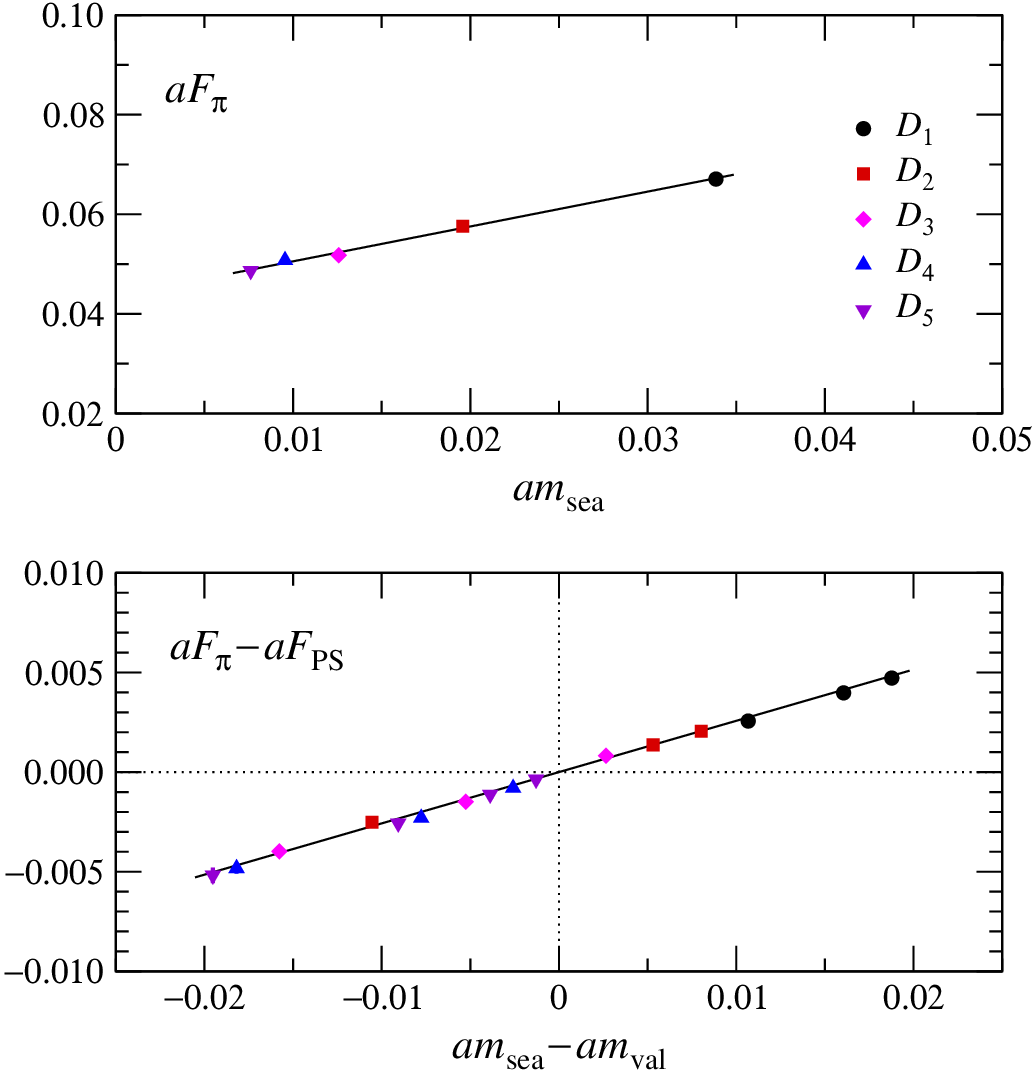}%
\vskip0.3cm
\figurecaption{%
Dependence of the bare pion decay constant 
$\Fpi=\left.\Fps\right|_{\mval=\msea}$ and of the difference 
$\Fpi-\Fps$ on the quark masses,
as determined on the $D$-series of lattices. 
The solid lines represent the global linear fit (6.3).
}
\vskip0.0cm
}
\endinsert

\subsection 6.2 Pseudo-scalar decay constant and vector meson mass

As can be seen from the tables in appendix C, the 
calculated values of $\Fps/\Mv$
are nearly independent of the quark masses. 
This comes a bit as a surprise, and could merely be
an accidental agreement in a limited range of masses,
since there does not appear to be any obvious 
physical connection between 
the pseudo-scalar decay constant and the vector meson mass.

The mass dependence of these two quantities is thus
practically the same and it suffices to consider 
one of them. Focusing on the decay constant,
a simple linear expression,
\equation{
  \Fps=b_0+b_1(\msea+\mval)+b_2\msea,
  \enum
}
turns out to fit the available data for $\Fps$ very well.
On the $D$-series of lattices, for example,
the fit matches the data within statistical
errors and the maximal relative deviation in the given
range of masses is only $1.6\%$ (see fig.~7). 

It is tempting to use these fits to extrapolate the 
decay constant to the chiral limit, but as already emphasized
in our previous paper [\ref{I}], such extrapolations are
difficult to justify and asymptotically
inconsistent with one-loop chiral perturbation theory.
On the other hand, the observed linearity of the pseudo-scalar decay
constant in the range of masses covered by the simulations
is striking and calls for a theoretical explanation.

\section 7. Concluding remarks

Numerical lattice QCD is currently in an interesting transition phase.
The valence approximation is now practically overcome, but
important physical effects of the light sea quarks,
such as the decay of the rho meson or the  
anomaly-driven mass splitting between the eta
and the pions,
still have not or only barely been studied directly.
Simulations at smaller quark masses and on 
larger lattices than reported here will probably be required for this.
Our experience however suggests that the prospects for
such simulations, using O($a$)-improved Wilson quarks,
are now quite good. 

So far the DD-HMC algorithm performed well and we did not 
run into any instabilities or other technical difficulties. 
As one moves to smaller quark masses and smaller lattice spacings,
there may be some room for 
further algorithmic improvements,
but the development of
variance-reduction methods is likely to be more rewarding
at this point,
particularly so if disconnected quark-line diagrams and 
multi-particle amplitudes are to be computed.

\vskip1.0ex
The numerical simulations were performed on PC
clusters at CERN, the Centro Enrico Fermi, the Institut f\"ur
Theoretische Physik der Universit\"at Bern (with a contribution from
the Schweizerischer Nationalfonds) and on a CRAY XT3 at the Swiss
National Super\-computing Centre (CSCS).
We are grateful to all these institutions for the continuous support
given to this project.

\appendix A. Statistical error analysis

In the physics analysis of the runs $A_1-A_3$, $B_1-B_4$ and 
$D_1-D_5$, we kept track of the statistical errors using
the jackknife method. In particular, 
any correlations among the errors of different observables
were always properly taken into account.
Here we summarize our conventions and briefly explain
the basic procedures that we used.

\subsection A.1 Jackknife samples

Let $A_{r}$, $r=1,\ldots,R$, be a set of 
primary stochastic observables and 
$a_{r,1},\ldots,a_{r,N}$ a sequence of $N$ measured values of
these. In lattice QCD the most common primary observables are the Wilson
loops and sums of products of quark propagators.
The jackknife method assumes that
the measured values are unbiased and statistically independent.
We shall thus take it for granted that the residual autocorrelations
are negligible in the cases of interest (see sect.~4).

The averages $\abar_r$ of the observables $A_r$
and the associated statistical error covariance $C_{rs}$
are given by
\equation{
   \abar_r={1\over N}\sum_{i=1}^Na_{r,i},
   \enum
   \next{2ex}
   C_{rs}={1\over N(N-1)}\sum_{i=1}^N
   \left(a_{r,i}-\abar_r\right)
   \left(a_{s,i}-\abar_s\right).
   \enum
}
If we introduce 
the jackknife samples
\equation{
   a^J_{r,i}=\abar_r+c_N\left(\abar_r-a_{r,i}\right),
   \qquad
   c_N=\left(N(N-1)\right)^{-1/2},
   \enum
}
an equivalent expression for the error matrix is
\equation{
   C_{rs}=\sum_{i=1}^N
   \left(a^J_{r,i}-\abar_r\right)
   \left(a^J_{s,i}-\abar_s\right).
   \enum
}
Note that our definition of the jackknife
samples slightly departs from
the standard conventions, where $c_N=1/(N-1)$. 
The modification is numerically insignificant in practice, but 
leads to some simplifications
when data from different simulations
are to be combined (see subsect.~A.3).

\subsection A.2 Error propagation

Apart from estimating the primary observables, one may be
interested in evaluating various functions $f(A_1,\ldots,A_R)$
of them,
which may involve fit procedures and 
other complicated operations. 
The standard stochastic estimate of such an observable is
\equation{
   \fbar=f(\abar_1,\ldots,\abar_R)
   \enum
}
and the associated series of jackknife estimates is defined by
\equation{
   f^J_i=f(\abar^J_{1,i},\ldots,\abar^J_{R,i}), 
   \qquad i=1,\ldots,N.
   \enum
}
A little algebra then shows that the expression
\equation{
  \sigma^2=
  \sum_{i=1}^N
  \left(f^J_i-\fbar\right)^2
  \enum
}
provides an estimate of the statistical variance of $\fbar$, which
coincides with the usual error propagation formula (the one 
that involves the
gradient of $f$) up to terms of order $1/N$. Similarly the 
error covariance of $f$ and any other function $g$ is obtained by 
summing $(f^J_i-\fbar)(g^J_i-\gbar)$
over the jackknife samples.

In practice the error formula (A.7) proves to be very convenient.
If an observable is a function of previously calculated
observables, for example, one can take advantage of the 
fact that the composition of functions is associative, i.e.~the 
jackknife series $f^J_i$ is simply obtained
by inserting the jackknife series of the arguments,
independently of whether these are primary or not.
The data analysis can thus proceed in steps, starting from the 
primary observables and progressing to more and more complicated
observables.

\subsection A.3 Combining data from different runs

Simulations of lattice QCD at different
sea-quark masses, lattice spacings, etc., can be assumed to 
be statistically independent. The statistical variance of any observable
that depends on data from several simulations is therefore the sum of the 
associated partial variances. 
This rule can easily be accommodated in the jackknife analysis
by embedding the jackknife series of the observables in extended 
series that include all simulations on which the 
observable depends.

The method is best explained by considering two
simulations, where $N_1$ measurements
of some observables $A_r$ are made in the first
and $N_2$ measurements of some other observables $B_s$
in the second. The associated jackknife
series $a^J_{r,1},\ldots,a^J_{r,N_1}$ and $b^J_{s,1},\ldots,b^J_{s,N_2}$
are then computed as before,
starting from the primary observables in each simulation.
Next they are embedded in extended series 
\equation{
   a^J_{r,1},\ldots,a^J_{r,N_1},
   \underbrace{\abar_r,\ldots,\abar_r}_{N_2\;{\rm elements}}
   \qquad\hbox{and}\qquad 
   \underbrace{\bbar_s,\ldots,\bbar_s}_{N_1\;{\rm elements}}
   b^J_{s,1},\ldots,b^J_{s,N_2}
   \enum
}
of length $N_1+N_2$ 
such that the first $N_1$ elements are occupied by the jackknife
series from the first simulation and the last $N_2$ elements by those from the 
second simulation. 

With this assignment, and if the extended
series are treated as ordinary jackknife series,
the correct error correlation matrix of the full set 
$A_1,\ldots,A_R,B_1,\ldots,B_S$
of observables is obtained.
Moreover, we may define the jackknife series of 
any observable $f(A_1,\ldots A_R,B_1,\ldots,B_S)$ in the standard
manner and compute its variance using eq.~(A.7).
The embedding trick thus allows the statistical errors
to be pro\-pa\-gated as if there were a single simulation.

\appendix B. Properties of the auxiliary function $\mib h(t;E,E')$

The symmetry properties
\equation{
  h(t;E,E')=h(T-t;E,E')=h(t;E',E)
  \enum
}
are an immediate consequence of the definition (5.9)
of the function $h(t;E,E')$. It is
also straightforward to verify that
\equation{
  h(t;E,E')=2\kern0.5pt
  \rme^{-{1\over2}(E+E')T}\cosh\left(\frac{1}{2}(E'-E)(T-2t)\right),
  \enum
}
and $h(t;E,E')$ is thus a convex function of $t$ which attains
its minimum at $t=\frac{1}{2}T$.

\subsection B.1 Product inequality

We now show that the inequality 
\equation{
  h(t;E,E'+E'')\leq h(t;E,E')h(t;0,E'')
  \enum
}
holds for all values of the arguments $t$, $E$, $E'$ and $E''$.
To this end, first note that  
\equation{
  \cosh(\alpha+\beta)\leq \cosh(\alpha+\beta)+\cosh(\alpha-\beta)=
  2\cosh\alpha\cosh\beta.
  \enum
}
Substituting 
$\alpha=\frac{1}{2}(E'-E)(T-2t)$ and 
$\beta=\frac{1}{2}E''(T-2t)$, this inequality becomes
\equation{
  \cosh\left(\frac{1}{2}(E'+E''-E)(T-2t)\right)
  \leq
  \noenum
  \next{2ex}
  \qquad2
  \cosh\left(\frac{1}{2}(E'-E)(T-2t)\right)
  \cosh\left(\frac{1}{2}E''(T-2t)\right),
  \enum
}
which is easily seen to coincide with (B.3) after inserting 
the representation (B.2).

\subsection B.2 Monotonicity property

If $t$ and $s$ are 
in the range $s<t\leq\frac{1}{2}T$, and if $M>0$,
it follows from eq.~(B.2) that the ratio
\equation{
   r={h(s;0,M)\over h(t;0,M)}
   \enum
}
is greater than $1$. A less obvious
statement is that the ratio increases monotonically
from $r=1$ to $r=\infty$ when $M$ goes from zero to infinity.

In order to show this, we insert eq.~(B.2) and work out the 
quotient
\equation{
  q={r-1\over r+1}=
  \tanh\left(\frac{1}{2}M(t-s)\right)
  \tanh\left(\frac{1}{2}M(T-t-s)\right).
  \enum
}
In the specified range of $t$ and $s$, the arguments of the hyperbolic
functions in this equation are non-negative and monotonically
growing with $M$. The quotient thus rises monotonically
from $0$ to $1$ when $M$ goes from zero to infinity,
which proves our claim, since $r$ and $q$ are monotonically related
to each other.

\appendix C. Tables of meson masses and decay constants 

The simulation results tabulated in this appendix were obtained
following the lines of sect.~5. In all cases, the $r$ quark
was taken to be a sea quark, i.e.~the
associated hopping parameter $\kappa_r$ was set to $\ksea$.
The hopping parameter $\kappa_s$ of the other quark, on the other
hand, ranged over $4$ or $5$ values, one of which being $\ksea$.

For each series of runs, we quote the quark mass sums $m_{rs}$,
the pseudo-scalar meson masses $\Mps$ and matrix elements $\Gps$,
and the vector meson masses $\Mv$, all given in lattice units 
(tables 5, 7 and 9).
Some combinations of these quantities are
printed in tables 6, 8 and 10. The errors given in brackets 
are statistical only.

If so desired, the quoted results can be 
converted to physical units by substituting the estimates
$0.0717(15)$, $0.0521(7)$ and $0.0784(10)$~fm for the
spacings of the $A$, $B$ and $D$ lattices [\ref{I}].
The quark mass sums $m_{rs}$, the matrix elements $\Gps$ and 
the decay constants $\Fps$ then also need to be renormalized,
\equation{
   m_{rs}\to\ZA\ZP^{-1}m_{rs},
   \qquad
   \Gps\to\ZP\Gps,
   \qquad
   \Fps\to\ZA\Fps,
   \enum
}
where $\ZA$ and $\ZP$ denote the (mass-independent) 
renormalization constants of the non-singlet axial current and density.
Moreover, in order to guarantee the O($a$) improvement of 
these quantities in the improved theory, the renormalization constants
must be modified according to
\equation{
   \ZX\to\ZX\bigl(1+\bXsea a\msea+\frac{1}{2}\bXval am_{rs}\bigr),
   \enum
}
with properly adjusted coefficients $\bXsea$ and $\bXval$
[\ref{OaImp},\ref{OaImpNonDeg}]
(the figures quoted in tables 9 and 10 include the contribution 
of the operator improvement term proportional to $\cA$ but not
the $1+\rmO(am)$ renormalization factors).

\vfill\eject

\pageinsert
\newdimen\digitwidth
\setbox0=\hbox{\rm 0}
\digitwidth=\wd0
\catcode`@=\active
\def@{\kern\digitwidth}
\tablecaption{Results for $m_{rs}$, $\Mps$, $\Gps$ and $\Mv$ 
(lattices $A_1-A_3$)} 
\vskip-1.5ex
$$\vbox{\settabs\+&%
                  xxxx&xx&
                  xxxxxxxx&xx&
                  xxxxxxxx&xx&
                  xxxxxxxxx&xx&
                  xxxxxxxxx&xx&
                  xxxxxxxxx&xx&
                  xxxxxxxxx&x\cr
\thicktablerule
\vskip1.0ex
                \+& \hfill Run \hfill
                 && \hfill $\kappa_r$\hfill
                 && \hfill $\kappa_s$\hfill
                 && \hfill $am_{rs}$\hfill
                 && \hfill $a\Mps$\hfill
                 && \hfill $a^2\Gps$\hfill
                 && \hfill $a\Mv$\hfill
                 &\cr
\vskip1.0ex
\thintablerule
\vskip1.5ex
  \+& \hfill $A_1$\hfill
  &&  \hfill $0.15750$\hfill
  &&  \hfill $0.15750$\hfill 
  &&  \hfill $0.0548(5)$\hfill
  &&  \hfill $0.2726(19)$\hfill
  &&  \hfill $0.0881(12)$\hfill
  &&  \hfill $0.389(4)$\hfill
  &\cr
\vskip0.3ex
  \+& \hfill $\;$\hfill
  &&  \hfill $$\hfill
  &&  \hfill $0.15800$\hfill 
  &&  \hfill $0.0472(6)$\hfill
  &&  \hfill $0.2536(19)$\hfill
  &&  \hfill $0.0859(12)$\hfill
  &&  \hfill $0.379(5)$\hfill
  &\cr
\vskip0.3ex
  \+& \hfill $\;$\hfill
  &&  \hfill $$\hfill
  &&  \hfill $0.15825$\hfill 
  &&  \hfill $0.0434(6)$\hfill
  &&  \hfill $0.2438(20)$\hfill
  &&  \hfill $0.0848(13)$\hfill
  &&  \hfill $0.373(5)$\hfill
  &\cr
\vskip0.3ex
  \+& \hfill $\;$\hfill
  &&  \hfill $$\hfill
  &&  \hfill $0.15835$\hfill 
  &&  \hfill $0.0419(6)$\hfill
  &&  \hfill $0.2398(21)$\hfill
  &&  \hfill $0.0844(13)$\hfill
  &&  \hfill $0.371(5)$\hfill
  &\cr
\vskip1.0ex
\thintablerule
\vskip1.5ex
  \+& \hfill $A_2$\hfill
  &&  \hfill $0.15800$\hfill
  &&  \hfill $0.15750$\hfill 
  &&  \hfill $0.0359(3)$\hfill
  &&  \hfill $0.2137(18)$\hfill
  &&  \hfill $0.0703(12)$\hfill
  &&  \hfill $0.344(3)$\hfill
  &\cr
\vskip0.3ex
  \+& \hfill $\;$\hfill
  &&  \hfill $$\hfill
  &&  \hfill $0.15800$\hfill 
  &&  \hfill $0.0285(3)$\hfill
  &&  \hfill $0.1913(19)$\hfill
  &&  \hfill $0.0682(13)$\hfill
  &&  \hfill $0.334(4)$\hfill
  &\cr
\vskip0.3ex
  \+& \hfill $\;$\hfill
  &&  \hfill $$\hfill
  &&  \hfill $0.15825$\hfill 
  &&  \hfill $0.0249(3)$\hfill
  &&  \hfill $0.1790(21)$\hfill
  &&  \hfill $0.0671(14)$\hfill
  &&  \hfill $0.329(5)$\hfill
  &\cr
\vskip0.3ex
  \+& \hfill $\;$\hfill
  &&  \hfill $$\hfill
  &&  \hfill $0.15835$\hfill 
  &&  \hfill $0.0235(3)$\hfill
  &&  \hfill $0.1738(22)$\hfill
  &&  \hfill $0.0666(15)$\hfill
  &&  \hfill $0.328(5)$\hfill
  &\cr
\vskip1.0ex
\thintablerule
\vskip1.5ex
  \+& \hfill $A_3$\hfill
  &&  \hfill $0.15825$\hfill
  &&  \hfill $0.15750$\hfill 
  &&  \hfill $0.0281(4)$\hfill
  &&  \hfill $0.185(3)@@$\hfill
  &&  \hfill $0.0617(19)$\hfill
  &&  \hfill $0.327(5)$\hfill
  &\cr
\vskip0.3ex
  \+& \hfill $\;$\hfill
  &&  \hfill $$\hfill
  &&  \hfill $0.15800$\hfill 
  &&  \hfill $0.0208(4)$\hfill
  &&  \hfill $0.160(3)@@$\hfill
  &&  \hfill $0.0599(22)$\hfill
  &&  \hfill $0.317(7)$\hfill
  &\cr
\vskip0.3ex
  \+& \hfill $\;$\hfill
  &&  \hfill $$\hfill
  &&  \hfill $0.15825$\hfill 
  &&  \hfill $0.0172(4)$\hfill
  &&  \hfill $0.147(4)@@$\hfill
  &&  \hfill $0.0593(23)$\hfill
  &&  \hfill $0.312(8)$\hfill
  &\cr
\vskip0.3ex
  \+& \hfill $\;$\hfill
  &&  \hfill $$\hfill
  &&  \hfill $0.15835$\hfill 
  &&  \hfill $0.0158(4)$\hfill
  &&  \hfill $0.141(4)@@$\hfill
  &&  \hfill $0.0592(24)$\hfill
  &&  \hfill $0.311(9)$\hfill
  &\cr
\vskip1.5ex
\thicktablerule
\vskip1.5ex
}
$$
\vfill
\endinsert

\pageinsert
\newdimen\digitwidth
\setbox0=\hbox{\rm 0}
\digitwidth=\wd0
\catcode`@=\active
\def@{\kern\digitwidth}
\tablecaption{Combinations of $m_{rs}$, $\Mps$, $\Gps$ and $\Mv$ 
(lattices $A_1-A_3$)} 
\vskip-1.5ex
$$\vbox{\settabs\+&%
                  xxxx&xx&
                  xxxxxxxx&xx&
                  xxxxxxxx&xx&
                  xxxxxxxxxx&xx&
                  xxxxxxxxxx&xx&
                  xxxxxxxxxx&x\cr
\thicktablerule
\vskip1.0ex
                \+& \hfill Run \hfill
                 && \hfill $\kappa_r$\hfill
                 && \hfill $\kappa_s$\hfill
                 && \hfill $a\Mps^2/m_{rs}$\hfill
                 && \hfill $a\Fps@$\hfill
                 && \hfill $\Fps/\Mv$\hfill
                 &\cr
\vskip1.0ex
\thintablerule
\vskip1.5ex
  \+& \hfill $A_1$\hfill
  &&  \hfill $0.15750$\hfill
  &&  \hfill $0.15750$\hfill 
  &&  \hfill $1.357(17)$\hfill
  &&  \hfill $0.0650(7)@$\hfill
  &&  \hfill $0.1669(21)$\hfill
  &\cr
\vskip0.3ex
  \+& \hfill $\;$\hfill
  &&  \hfill $$\hfill
  &&  \hfill $0.15800$\hfill 
  &&  \hfill $1.363(19)$\hfill
  &&  \hfill $0.0630(7)@$\hfill
  &&  \hfill $0.1664(24)$\hfill
  &\cr
\vskip0.3ex
  \+& \hfill $\;$\hfill
  &&  \hfill $$\hfill
  &&  \hfill $0.15825$\hfill 
  &&  \hfill $1.369(21)$\hfill
  &&  \hfill $0.0619(8)@$\hfill
  &&  \hfill $0.166(3)@@$\hfill
  &\cr
\vskip0.3ex
  \+& \hfill $\;$\hfill
  &&  \hfill $$\hfill
  &&  \hfill $0.15835$\hfill 
  &&  \hfill $1.372(22)$\hfill
  &&  \hfill $0.0615(8)@$\hfill
  &&  \hfill $0.166(3)@@$\hfill
  &\cr
\vskip1.0ex
\thintablerule
\vskip1.5ex
  \+& \hfill $A_2$\hfill
  &&  \hfill $0.15800$\hfill
  &&  \hfill $0.15750$\hfill 
  &&  \hfill $1.272(20)$\hfill
  &&  \hfill $0.0553(7)@$\hfill
  &&  \hfill $0.161(3)@@$\hfill
  &\cr
\vskip0.3ex
  \+& \hfill $\;$\hfill
  &&  \hfill $$\hfill
  &&  \hfill $0.15800$\hfill 
  &&  \hfill $1.282(25)$\hfill
  &&  \hfill $0.0532(7)@$\hfill
  &&  \hfill $0.159(3)@@$\hfill
  &\cr
\vskip0.3ex
  \+& \hfill $\;$\hfill
  &&  \hfill $$\hfill
  &&  \hfill $0.15825$\hfill 
  &&  \hfill $1.29(3)@@$\hfill
  &&  \hfill $0.0522(8)@$\hfill
  &&  \hfill $0.159(3)@@$\hfill
  &\cr
\vskip0.3ex
  \+& \hfill $\;$\hfill
  &&  \hfill $$\hfill
  &&  \hfill $0.15835$\hfill 
  &&  \hfill $1.29(3)@@$\hfill
  &&  \hfill $0.0518(8)@$\hfill
  &&  \hfill $0.158(4)@@$\hfill
  &\cr
\vskip1.0ex
\thintablerule
\vskip1.5ex
  \+& \hfill $A_3$\hfill
  &&  \hfill $0.15825$\hfill
  &&  \hfill $0.15750$\hfill 
  &&  \hfill $1.22(4)@@$\hfill
  &&  \hfill $0.0505(8)@$\hfill
  &&  \hfill $0.154(3)@@$\hfill
  &\cr
\vskip0.3ex
  \+& \hfill $\;$\hfill
  &&  \hfill $$\hfill
  &&  \hfill $0.15800$\hfill 
  &&  \hfill $1.23(5)@@$\hfill
  &&  \hfill $0.0486(10)$\hfill
  &&  \hfill $0.153(4)@@$\hfill
  &\cr
\vskip0.3ex
  \+& \hfill $\;$\hfill
  &&  \hfill $$\hfill
  &&  \hfill $0.15825$\hfill 
  &&  \hfill $1.25(6)@@$\hfill
  &&  \hfill $0.0474(11)$\hfill
  &&  \hfill $0.152(5)@@$\hfill
  &\cr
\vskip0.3ex
  \+& \hfill $\;$\hfill
  &&  \hfill $$\hfill
  &&  \hfill $0.15835$\hfill 
  &&  \hfill $1.26(7)@@$\hfill
  &&  \hfill $0.0469(12)$\hfill
  &&  \hfill $0.151(6)@@$\hfill
  &\cr
\vskip1.5ex
\thicktablerule
\vskip1.5ex
}
$$
\vfill
\endinsert

\pageinsert
\newdimen\digitwidth
\setbox0=\hbox{\rm 0}
\digitwidth=\wd0
\catcode`@=\active
\def@{\kern\digitwidth}
\tablecaption{Results for $m_{rs}$, $\Mps$, $\Gps$ and $\Mv$ 
(lattices $B_1-B_4$)} 
\vskip-1.5ex
$$\vbox{\settabs\+&%
                  xxxx&xx&
                  xxxxxxxx&x&
                  xxxxxxxx&xx&
                  xxxxxxxxx&xxx&
                  xxxxxxxxx&xx&
                  xxxxxxxxx&xx&
                  xxxxxxxxx&x\cr
\thicktablerule
\vskip1.0ex
                \+& \hfill Run \hfill
                 && \hfill $\kappa_r$\hfill
                 && \hfill $\kappa_s$\hfill
                 && \hfill $am_{rs}$\hfill
                 && \hfill $a\Mps$\hfill
                 && \hfill $a^2\Gps$\hfill
                 && \hfill $a\Mv$\hfill
                 &\cr
\vskip1.0ex
\thintablerule
\vskip1.5ex
  \+& \hfill $B_1$\hfill
  &&  \hfill $0.15410$\hfill
  &&  \hfill $0.15410$\hfill 
  &&  \hfill $0.03889(18)$\hfill
  &&  \hfill $0.1958(9)@$\hfill
  &&  \hfill $0.0453(5)$\hfill
  &&  \hfill $0.2896(17)$\hfill
  &\cr
\vskip0.3ex
  \+& \hfill $\;$\hfill
  &&  \hfill $$\hfill
  &&  \hfill $0.15425$\hfill 
  &&  \hfill $0.03631(18)$\hfill
  &&  \hfill $0.1892(9)@$\hfill
  &&  \hfill $0.0447(5)$\hfill
  &&  \hfill $0.2858(17)$\hfill
  &\cr
\vskip0.3ex
  \+& \hfill $\;$\hfill
  &&  \hfill $$\hfill
  &&  \hfill $0.15440$\hfill 
  &&  \hfill $0.03375(18)$\hfill
  &&  \hfill $0.1824(10)$\hfill
  &&  \hfill $0.0441(6)$\hfill
  &&  \hfill $0.2820(18)$\hfill
  &\cr
\vskip0.3ex
  \+& \hfill $\;$\hfill
  &&  \hfill $$\hfill
  &&  \hfill $0.15455$\hfill 
  &&  \hfill $0.03120(18)$\hfill
  &&  \hfill $0.1754(10)$\hfill
  &&  \hfill $0.0435(6)$\hfill
  &&  \hfill $0.2782(19)$\hfill
  &\cr
\vskip1.0ex
\thintablerule
\vskip1.5ex
  \+& \hfill $B_2$\hfill
  &&  \hfill $0.15440$\hfill
  &&  \hfill $0.15410$\hfill 
  &&  \hfill $0.02696(13)$\hfill
  &&  \hfill $0.1619(11)$\hfill
  &&  \hfill $0.0384(7)$\hfill
  &&  \hfill $0.2518(21)$\hfill
  &\cr
\vskip0.3ex
  \+& \hfill $\;$\hfill
  &&  \hfill $$\hfill
  &&  \hfill $0.15425$\hfill 
  &&  \hfill $0.02440(14)$\hfill
  &&  \hfill $0.1546(12)$\hfill
  &&  \hfill $0.0379(7)$\hfill
  &&  \hfill $0.2475(22)$\hfill
  &\cr
\vskip0.3ex
  \+& \hfill $\;$\hfill
  &&  \hfill $$\hfill
  &&  \hfill $0.15440$\hfill 
  &&  \hfill $0.02187(14)$\hfill
  &&  \hfill $0.1470(12)$\hfill
  &&  \hfill $0.0374(7)$\hfill
  &&  \hfill $0.2432(24)$\hfill
  &\cr
\vskip0.3ex
  \+& \hfill $\;$\hfill
  &&  \hfill $$\hfill
  &&  \hfill $0.15455$\hfill 
  &&  \hfill $0.01935(14)$\hfill
  &&  \hfill $0.1391(13)$\hfill
  &&  \hfill $0.0370(8)$\hfill
  &&  \hfill $0.239(3)@@$\hfill
  &\cr
\vskip1.0ex
\thintablerule
\vskip1.5ex
  \+& \hfill $B_3$\hfill
  &&  \hfill $0.15455$\hfill
  &&  \hfill $0.15410$\hfill 
  &&  \hfill $0.02185(12)$\hfill
  &&  \hfill $0.1416(12)$\hfill
  &&  \hfill $0.0333(7)$\hfill
  &&  \hfill $0.2418(24)$\hfill
  &\cr
\vskip0.3ex
  \+& \hfill $\;$\hfill
  &&  \hfill $$\hfill
  &&  \hfill $0.15425$\hfill 
  &&  \hfill $0.01927(12)$\hfill
  &&  \hfill $0.1329(13)$\hfill
  &&  \hfill $0.0326(7)$\hfill
  &&  \hfill $0.238(3)@@$\hfill
  &\cr
\vskip0.3ex
  \+& \hfill $\;$\hfill
  &&  \hfill $$\hfill
  &&  \hfill $0.15440$\hfill 
  &&  \hfill $0.01668(12)$\hfill
  &&  \hfill $0.1235(14)$\hfill
  &&  \hfill $0.0318(7)$\hfill
  &&  \hfill $0.233(3)@@$\hfill
  &\cr
\vskip0.3ex
  \+& \hfill $\;$\hfill
  &&  \hfill $$\hfill
  &&  \hfill $0.15455$\hfill 
  &&  \hfill $0.01409(13)$\hfill
  &&  \hfill $0.1132(15)$\hfill
  &&  \hfill $0.0310(8)$\hfill
  &&  \hfill $0.230(3)@@$\hfill
  &\cr
\vskip1.0ex
\thintablerule
\vskip1.5ex
  \+& \hfill $B_4$\hfill
  &&  \hfill $0.15462$\hfill
  &&  \hfill $0.15410$\hfill 
  &&  \hfill $0.02029(16)$\hfill
  &&  \hfill $0.1328(10)$\hfill
  &&  \hfill $0.0317(6)$\hfill
  &&  \hfill $0.237(3)@@$\hfill
  &\cr
\vskip0.3ex
  \+& \hfill $\;$\hfill
  &&  \hfill $$\hfill
  &&  \hfill $0.15425$\hfill 
  &&  \hfill $0.01774(16)$\hfill
  &&  \hfill $0.1242(11)$\hfill
  &&  \hfill $0.0312(6)$\hfill
  &&  \hfill $0.233(3)@@$\hfill
  &\cr
\vskip0.3ex
  \+& \hfill $\;$\hfill
  &&  \hfill $$\hfill
  &&  \hfill $0.15440$\hfill 
  &&  \hfill $0.01521(17)$\hfill
  &&  \hfill $0.1151(12)$\hfill
  &&  \hfill $0.0307(6)$\hfill
  &&  \hfill $0.229(3)@@$\hfill 
 &\cr
\vskip0.3ex
  \+& \hfill $\;$\hfill
  &&  \hfill $$\hfill
  &&  \hfill $0.15455$\hfill 
  &&  \hfill $0.01269(17)$\hfill
  &&  \hfill $0.1055(14)$\hfill
  &&  \hfill $0.0302(7)$\hfill
  &&  \hfill $0.224(4)@@$\hfill
  &\cr
\vskip0.3ex
  \+& \hfill $\;$\hfill
  &&  \hfill $$\hfill
  &&  \hfill $0.15462$\hfill 
  &&  \hfill $0.01151(17)$\hfill
  &&  \hfill $0.1008(15)$\hfill
  &&  \hfill $0.0300(8)$\hfill
  &&  \hfill $0.223(4)@@$\hfill
  &\cr
\vskip1.5ex
\thicktablerule
\vskip1.5ex
}
$$
\vfill
\endinsert

\pageinsert
\newdimen\digitwidth
\setbox0=\hbox{\rm 0}
\digitwidth=\wd0
\catcode`@=\active
\def@{\kern\digitwidth}
\tablecaption{Combinations of $m_{rs}$, $\Mps$, $\Gps$ and $\Mv$ 
(lattices $B_1-B_4$)} 
\vskip-1.5ex
$$\vbox{\settabs\+&%
                  xxxx&xx&
                  xxxxxxxx&xx&
                  xxxxxxxx&xx&
                  xxxxxxxxxx&xx&
                  xxxxxxxxxx&xx&
                  xxxxxxxxxx&x\cr
\thicktablerule
\vskip1.0ex
                \+& \hfill Run \hfill
                 && \hfill $\kappa_r$\hfill
                 && \hfill $\kappa_s$\hfill
                 && \hfill $a\Mps^2/m_{rs}$\hfill
                 && \hfill $a\Fps@$\hfill
                 && \hfill $\Fps/\Mv$\hfill
                 &\cr
\vskip1.0ex
\thintablerule
\vskip1.5ex
  \+& \hfill $B_1$\hfill
  &&  \hfill $0.15410$\hfill
  &&  \hfill $0.15410$\hfill 
  &&  \hfill $0.986(7)@$\hfill
  &&  \hfill $0.0460(4)$\hfill
  &&  \hfill $0.1587(17)$\hfill
  &\cr
\vskip0.3ex
  \+& \hfill $\;$\hfill
  &&  \hfill $$\hfill
  &&  \hfill $0.15425$\hfill 
  &&  \hfill $0.986(8)@$\hfill
  &&  \hfill $0.0453(4)$\hfill
  &&  \hfill $0.1586(18)$\hfill
  &\cr
\vskip0.3ex
  \+& \hfill $\;$\hfill
  &&  \hfill $$\hfill
  &&  \hfill $0.15440$\hfill 
  &&  \hfill $0.986(8)@$\hfill
  &&  \hfill $0.0447(4)$\hfill
  &&  \hfill $0.1585(19)$\hfill
  &\cr
\vskip0.3ex
  \+& \hfill $\;$\hfill
  &&  \hfill $$\hfill
  &&  \hfill $0.15455$\hfill 
  &&  \hfill $0.986(9)@$\hfill
  &&  \hfill $0.0441(5)$\hfill
  &&  \hfill $0.1584(20)$\hfill
  &\cr
\vskip1.0ex
\thintablerule
\vskip1.5ex
  \+& \hfill $B_2$\hfill
  &&  \hfill $0.15440$\hfill
  &&  \hfill $0.15410$\hfill 
  &&  \hfill $0.973(14)$\hfill
  &&  \hfill $0.0395(4)$\hfill
  &&  \hfill $0.1567(19)$\hfill
  &\cr
\vskip0.3ex
  \+& \hfill $\;$\hfill
  &&  \hfill $$\hfill
  &&  \hfill $0.15425$\hfill 
  &&  \hfill $0.979(15)$\hfill
  &&  \hfill $0.0387(4)$\hfill
  &&  \hfill $0.1562(20)$\hfill
  &\cr
\vskip0.3ex
  \+& \hfill $\;$\hfill
  &&  \hfill $$\hfill
  &&  \hfill $0.15440$\hfill 
  &&  \hfill $0.988(17)$\hfill
  &&  \hfill $0.0379(4)$\hfill
  &&  \hfill $0.1556(21)$\hfill
  &\cr
\vskip0.3ex
  \+& \hfill $\;$\hfill
  &&  \hfill $$\hfill
  &&  \hfill $0.15455$\hfill 
  &&  \hfill $1.000(19)$\hfill
  &&  \hfill $0.0370(4)$\hfill
  &&  \hfill $0.1549(23)$\hfill
  &\cr
\vskip1.0ex
\thintablerule
\vskip1.5ex
  \+& \hfill $B_3$\hfill
  &&  \hfill $0.15455$\hfill
  &&  \hfill $0.15410$\hfill 
  &&  \hfill $0.918(15)$\hfill
  &&  \hfill $0.0363(3)$\hfill
  &&  \hfill $0.1502(21)$\hfill
  &\cr
\vskip0.3ex
  \+& \hfill $\;$\hfill
  &&  \hfill $$\hfill
  &&  \hfill $0.15425$\hfill 
  &&  \hfill $0.916(17)$\hfill
  &&  \hfill $0.0356(4)$\hfill
  &&  \hfill $0.1497(23)$\hfill
  &\cr
\vskip0.3ex
  \+& \hfill $\;$\hfill
  &&  \hfill $$\hfill
  &&  \hfill $0.15440$\hfill 
  &&  \hfill $0.914(19)$\hfill
  &&  \hfill $0.0348(4)$\hfill
  &&  \hfill $0.149(3)@@$\hfill
  &\cr
\vskip0.3ex
  \+& \hfill $\;$\hfill
  &&  \hfill $$\hfill
  &&  \hfill $0.15455$\hfill 
  &&  \hfill $0.910(22)$\hfill
  &&  \hfill $0.0340(4)$\hfill
  &&  \hfill $0.148(3)@@$\hfill
  &\cr
\vskip1.0ex
\thintablerule
\vskip1.5ex
  \+& \hfill $B_4$\hfill
  &&  \hfill $0.15462$\hfill
  &&  \hfill $0.15410$\hfill 
  &&  \hfill $0.869(13)$\hfill
  &&  \hfill $0.0365(4)$\hfill
  &&  \hfill $0.154(3)@@$\hfill
  &\cr
\vskip0.3ex
  \+& \hfill $\;$\hfill
  &&  \hfill $$\hfill
  &&  \hfill $0.15425$\hfill 
  &&  \hfill $0.869(15)$\hfill
  &&  \hfill $0.0359(4)$\hfill
  &&  \hfill $0.154(3)@@$\hfill
  &\cr
\vskip0.3ex
  \+& \hfill $\;$\hfill
  &&  \hfill $$\hfill
  &&  \hfill $0.15440$\hfill 
  &&  \hfill $0.871(18)$\hfill
  &&  \hfill $0.0352(5)$\hfill
  &&  \hfill $0.154(3)@@$\hfill
  &\cr
\vskip0.3ex
  \+& \hfill $\;$\hfill
  &&  \hfill $$\hfill
  &&  \hfill $0.15455$\hfill 
  &&  \hfill $0.878(23)$\hfill
  &&  \hfill $0.0344(5)$\hfill
  &&  \hfill $0.153(4)@@$\hfill
  &\cr
\vskip0.3ex
  \+& \hfill $\;$\hfill
  &&  \hfill $$\hfill
  &&  \hfill $0.15462$\hfill 
  &&  \hfill $0.88(3)@@$\hfill
  &&  \hfill $0.0340(6)$\hfill
  &&  \hfill $0.153(4)@@$\hfill
  &\cr
\vskip1.5ex
\thicktablerule
\vskip1.5ex
}
$$
\vfill
\endinsert

\pageinsert
\newdimen\digitwidth
\setbox0=\hbox{\rm 0}
\digitwidth=\wd0
\catcode`@=\active
\def@{\kern\digitwidth}
\tablecaption{Results for $m_{rs}$, $\Mps$, $\Gps$ and $\Mv$ 
(lattices $D_1-D_5$)} 
\vskip-1.5ex
$$\vbox{\settabs\+&%
                  xxxx&xx&
                  xxxxxxxx&x&
                  xxxxxxxx&xx&
                  xxxxxxxxx&xxx&
                  xxxxxxxxx&xx&
                  xxxxxxxxx&xx&
                  xxxxxxxxx&x\cr
\thicktablerule
\vskip1.0ex
                \+& \hfill Run \hfill
                 && \hfill $\kappa_r$\hfill
                 && \hfill $\kappa_s$\hfill
                 && \hfill $am_{rs}$\hfill
                 && \hfill $a\Mps$\hfill
                 && \hfill $a^2\Gps$\hfill
                 && \hfill $a\Mv@@$\hfill
                 &\cr
\vskip1.0ex
\thintablerule
\vskip1.5ex
  \+& \hfill $D_1$\hfill
  &&  \hfill $0.13550$\hfill
  &&  \hfill $0.13550$\hfill 
  &&  \hfill $0.06771(21)$\hfill
  &&  \hfill $0.3286(10)$\hfill
  &&  \hfill $0.1069(15)$\hfill
  &&  \hfill $0.464(3)@@$\hfill
  &\cr
\vskip0.3ex
  \+& \hfill $\;$\hfill
  &&  \hfill $$\hfill
  &&  \hfill $0.13590$\hfill 
  &&  \hfill $0.05704(22)$\hfill
  &&  \hfill $0.3017(10)$\hfill
  &&  \hfill $0.1030(15)$\hfill
  &&  \hfill $0.447(3)@@$\hfill
  &\cr
\vskip0.3ex
  \+& \hfill $\;$\hfill
  &&  \hfill $$\hfill
  &&  \hfill $0.13610$\hfill 
  &&  \hfill $0.05165(23)$\hfill
  &&  \hfill $0.2873(11)$\hfill
  &&  \hfill $0.1008(15)$\hfill
  &&  \hfill $0.438(3)@@$\hfill
  &\cr
\vskip0.3ex
  \+& \hfill $\;$\hfill
  &&  \hfill $$\hfill
  &&  \hfill $0.13620$\hfill 
  &&  \hfill $0.04893(24)$\hfill
  &&  \hfill $0.2799(12)$\hfill
  &&  \hfill $0.0998(15)$\hfill
  &&  \hfill $0.434(4)@@$\hfill
  &\cr
\vskip1.0ex
\thintablerule
\vskip1.5ex
  \+& \hfill $D_2$\hfill
  &&  \hfill $0.13590$\hfill
  &&  \hfill $0.13550$\hfill 
  &&  \hfill $0.04968(13)$\hfill
  &&  \hfill $0.2758(8)@$\hfill
  &&  \hfill $0.0920(11)$\hfill
  &&  \hfill $0.4173(24)$\hfill
  &\cr
\vskip0.3ex
  \+& \hfill $\;$\hfill
  &&  \hfill $$\hfill
  &&  \hfill $0.13590$\hfill 
  &&  \hfill $0.03914(14)$\hfill
  &&  \hfill $0.2461(9)@$\hfill
  &&  \hfill $0.0891(11)$\hfill
  &&  \hfill $0.401(3)@@$\hfill
  &\cr
\vskip0.3ex
  \+& \hfill $\;$\hfill
  &&  \hfill $$\hfill
  &&  \hfill $0.13610$\hfill 
  &&  \hfill $0.03383(14)$\hfill
  &&  \hfill $0.2301(9)@$\hfill
  &&  \hfill $0.0880(12)$\hfill
  &&  \hfill $0.394(4)@@$\hfill
  &\cr
\vskip0.3ex
  \+& \hfill $\;$\hfill
  &&  \hfill $$\hfill
  &&  \hfill $0.13620$\hfill 
  &&  \hfill $0.03112(15)$\hfill
  &&  \hfill $0.2218(10)$\hfill
  &&  \hfill $0.0878(13)$\hfill
  &&  \hfill $0.390(4)@@$\hfill
  &\cr
\vskip1.0ex
\thintablerule
\vskip1.5ex
  \+& \hfill $D_3$\hfill
  &&  \hfill $0.13610$\hfill
  &&  \hfill $0.13550$\hfill 
  &&  \hfill $0.04092(14)$\hfill
  &&  \hfill $0.2440(10)$\hfill
  &&  \hfill $0.0811(12)$\hfill
  &&  \hfill $0.382(3)@@$\hfill
  &\cr
\vskip0.3ex
  \+& \hfill $\;$\hfill
  &&  \hfill $$\hfill
  &&  \hfill $0.13590$\hfill 
  &&  \hfill $0.03041(14)$\hfill
  &&  \hfill $0.2110(11)$\hfill
  &&  \hfill $0.0780(13)$\hfill
  &&  \hfill $0.363(4)@@$\hfill
  &\cr
\vskip0.3ex
  \+& \hfill $\;$\hfill
  &&  \hfill $$\hfill
  &&  \hfill $0.13610$\hfill 
  &&  \hfill $0.02514(15)$\hfill
  &&  \hfill $0.1929(12)$\hfill
  &&  \hfill $0.0766(14)$\hfill
  &&  \hfill $0.354(5)@@$\hfill
  &\cr
\vskip0.3ex
  \+& \hfill $\;$\hfill
  &&  \hfill $$\hfill
  &&  \hfill $0.13620$\hfill 
  &&  \hfill $0.02249(15)$\hfill
  &&  \hfill $0.1832(13)$\hfill
  &&  \hfill $0.0760(15)$\hfill
  &&  \hfill $0.349(5)@@$\hfill
  &\cr
\vskip1.0ex
\thintablerule
\vskip1.5ex
  \+& \hfill $D_4$\hfill
  &&  \hfill $0.13620$\hfill
  &&  \hfill $0.13550$\hfill 
  &&  \hfill $0.03728(14)$\hfill
  &&  \hfill $0.2335(11)$\hfill
  &&  \hfill $0.0813(11)$\hfill
  &&  \hfill $0.374(4)@@$\hfill
  &\cr
\vskip0.3ex
  \+& \hfill $\;$\hfill
  &&  \hfill $$\hfill
  &&  \hfill $0.13590$\hfill 
  &&  \hfill $0.02686(15)$\hfill
  &&  \hfill $0.1993(12)$\hfill
  &&  \hfill $0.0785(12)$\hfill
  &&  \hfill $0.356(4)@@$\hfill
  &\cr
\vskip0.3ex
  \+& \hfill $\;$\hfill
  &&  \hfill $$\hfill
  &&  \hfill $0.13610$\hfill 
  &&  \hfill $0.02168(15)$\hfill
  &&  \hfill $0.1800(13)$\hfill
  &&  \hfill $0.0771(13)$\hfill
  &&  \hfill $0.348(5)@@$\hfill
  &\cr
\vskip0.3ex
  \+& \hfill $\;$\hfill
  &&  \hfill $$\hfill
  &&  \hfill $0.13620$\hfill 
  &&  \hfill $0.01909(15)$\hfill
  &&  \hfill $0.1695(14)$\hfill
  &&  \hfill $0.0765(13)$\hfill
  &&  \hfill $0.345(6)@@$\hfill
  &\cr
\vskip1.0ex
\thintablerule
\vskip1.5ex
  \+& \hfill $D_5$\hfill
  &&  \hfill $0.13625$\hfill
  &&  \hfill $0.13550$\hfill 
  &&  \hfill $0.03474(13)$\hfill
  &&  \hfill $0.2249(11)$\hfill
  &&  \hfill $0.0784(13)$\hfill
  &&  \hfill $0.376(5)@@$\hfill
  &\cr
\vskip0.3ex
  \+& \hfill $\;$\hfill
  &&  \hfill $$\hfill
  &&  \hfill $0.13590$\hfill 
  &&  \hfill $0.02428(13)$\hfill
  &&  \hfill $0.1881(11)$\hfill
  &&  \hfill $0.0747(14)$\hfill
  &&  \hfill $0.359(6)@@$\hfill
  &\cr
\vskip0.3ex
  \+& \hfill $\;$\hfill
  &&  \hfill $$\hfill
  &&  \hfill $0.13610$\hfill 
  &&  \hfill $0.01910(13)$\hfill
  &&  \hfill $0.1672(13)$\hfill
  &&  \hfill $0.0729(15)$\hfill
  &&  \hfill $0.350(7)@@$\hfill
  &\cr
\vskip0.3ex
  \+& \hfill $\;$\hfill
  &&  \hfill $$\hfill
  &&  \hfill $0.13620$\hfill 
  &&  \hfill $0.01651(14)$\hfill
  &&  \hfill $0.1559(14)$\hfill
  &&  \hfill $0.0722(16)$\hfill
  &&  \hfill $0.346(8)@@$\hfill
  &\cr
\vskip0.3ex
  \+& \hfill $\;$\hfill
  &&  \hfill $$\hfill
  &&  \hfill $0.13625$\hfill 
  &&  \hfill $0.01522(14)$\hfill
  &&  \hfill $0.1499(15)$\hfill
  &&  \hfill $0.0719(17)$\hfill
  &&  \hfill $0.344(9)@@$\hfill
  &\cr
\vskip1.5ex
\thicktablerule
\vskip1.5ex
}
$$
\vfill
\endinsert

\pageinsert
\newdimen\digitwidth
\setbox0=\hbox{\rm 0}
\digitwidth=\wd0
\catcode`@=\active
\def@{\kern\digitwidth}
\tablecaption{Combinations of $m_{rs}$, $\Mps$, $\Gps$ and $\Mv$ 
(lattices $D_1-D_5$)} 
\vskip-1.5ex
$$\vbox{\settabs\+&%
                  xxxx&xx&
                  xxxxxxxx&xx&
                  xxxxxxxx&xx&
                  xxxxxxxxxx&xx&
                  xxxxxxxxxx&xx&
                  xxxxxxxxxx&x\cr
\thicktablerule
\vskip1.0ex
                \+& \hfill Run \hfill
                 && \hfill $\kappa_r$\hfill
                 && \hfill $\kappa_s$\hfill
                 && \hfill $a\Mps^2/m_{rs}$\hfill
                 && \hfill $a\Fps@$\hfill
                 && \hfill $\Fps/\Mv$\hfill
                 &\cr
\vskip1.0ex
\thintablerule
\vskip1.5ex
  \+& \hfill $D_1$\hfill
  &&  \hfill $0.13550$\hfill
  &&  \hfill $0.13550$\hfill 
  &&  \hfill $1.594(9)@$\hfill
  &&  \hfill $0.0671(9)$\hfill
  &&  \hfill $0.1445(20)$\hfill
  &\cr
\vskip0.3ex
  \+& \hfill $\;$\hfill
  &&  \hfill $$\hfill
  &&  \hfill $0.13590$\hfill 
  &&  \hfill $1.596(10)$\hfill
  &&  \hfill $0.0645(9)$\hfill
  &&  \hfill $0.1444(21)$\hfill
  &\cr
\vskip0.3ex
  \+& \hfill $\;$\hfill
  &&  \hfill $$\hfill
  &&  \hfill $0.13610$\hfill 
  &&  \hfill $1.598(11)$\hfill
  &&  \hfill $0.0631(9)$\hfill
  &&  \hfill $0.1440(23)$\hfill
  &\cr
\vskip0.3ex
  \+& \hfill $\;$\hfill
  &&  \hfill $$\hfill
  &&  \hfill $0.13620$\hfill 
  &&  \hfill $1.601(12)$\hfill
  &&  \hfill $0.0624(9)$\hfill
  &&  \hfill $0.1438(23)$\hfill
  &\cr
\vskip1.0ex
\thintablerule
\vskip1.5ex
  \+& \hfill $D_2$\hfill
  &&  \hfill $0.13590$\hfill
  &&  \hfill $0.13550$\hfill 
  &&  \hfill $1.531(9)@$\hfill
  &&  \hfill $0.0601(6)$\hfill
  &&  \hfill $0.1441(17)$\hfill
  &\cr
\vskip0.3ex
  \+& \hfill $\;$\hfill
  &&  \hfill $$\hfill
  &&  \hfill $0.13590$\hfill 
  &&  \hfill $1.547(10)$\hfill
  &&  \hfill $0.0576(7)$\hfill
  &&  \hfill $0.1435(20)$\hfill
  &\cr
\vskip0.3ex
  \+& \hfill $\;$\hfill
  &&  \hfill $$\hfill
  &&  \hfill $0.13610$\hfill 
  &&  \hfill $1.565(12)$\hfill
  &&  \hfill $0.0562(7)$\hfill
  &&  \hfill $0.1428(22)$\hfill
  &\cr
\vskip0.3ex
  \+& \hfill $\;$\hfill
  &&  \hfill $$\hfill
  &&  \hfill $0.13620$\hfill 
  &&  \hfill $1.581(14)$\hfill
  &&  \hfill $0.0556(7)$\hfill
  &&  \hfill $0.1424(24)$\hfill
  &\cr
\vskip1.0ex
\thintablerule
\vskip1.5ex
  \+& \hfill $D_3$\hfill
  &&  \hfill $0.13610$\hfill
  &&  \hfill $0.13550$\hfill 
  &&  \hfill $1.454(11)$\hfill
  &&  \hfill $0.0558(6)$\hfill
  &&  \hfill $0.1461(20)$\hfill
  &\cr
\vskip0.3ex
  \+& \hfill $\;$\hfill
  &&  \hfill $$\hfill
  &&  \hfill $0.13590$\hfill 
  &&  \hfill $1.465(14)$\hfill
  &&  \hfill $0.0533(7)$\hfill
  &&  \hfill $0.1467(23)$\hfill
  &\cr
\vskip0.3ex
  \+& \hfill $\;$\hfill
  &&  \hfill $$\hfill
  &&  \hfill $0.13610$\hfill 
  &&  \hfill $1.480(17)$\hfill
  &&  \hfill $0.0518(7)$\hfill
  &&  \hfill $0.146(3)@@$\hfill
  &\cr
\vskip0.3ex
  \+& \hfill $\;$\hfill
  &&  \hfill $$\hfill
  &&  \hfill $0.13620$\hfill 
  &&  \hfill $1.492(19)$\hfill
  &&  \hfill $0.0510(8)$\hfill
  &&  \hfill $0.146(3)@@$\hfill
  &\cr
\vskip1.0ex
\thintablerule
\vskip1.5ex
  \+& \hfill $D_4$\hfill
  &&  \hfill $0.13620$\hfill
  &&  \hfill $0.13550$\hfill 
  &&  \hfill $1.462(13)$\hfill
  &&  \hfill $0.0556(6)$\hfill
  &&  \hfill $0.1487(22)$\hfill
  &\cr
\vskip0.3ex
  \+& \hfill $\;$\hfill
  &&  \hfill $$\hfill
  &&  \hfill $0.13590$\hfill 
  &&  \hfill $1.478(17)$\hfill
  &&  \hfill $0.0531(7)$\hfill
  &&  \hfill $0.149(3)@@$\hfill
  &\cr
\vskip0.3ex
  \+& \hfill $\;$\hfill
  &&  \hfill $$\hfill
  &&  \hfill $0.13610$\hfill 
  &&  \hfill $1.494(20)$\hfill
  &&  \hfill $0.0516(7)$\hfill
  &&  \hfill $0.148(3)@@$\hfill
  &\cr
\vskip0.3ex
  \+& \hfill $\;$\hfill
  &&  \hfill $$\hfill
  &&  \hfill $0.13620$\hfill 
  &&  \hfill $1.505(23)$\hfill
  &&  \hfill $0.0508(7)$\hfill
  &&  \hfill $0.147(3)@@$\hfill
  &\cr
\vskip1.0ex
\thintablerule
\vskip1.5ex
  \+& \hfill $D_5$\hfill
  &&  \hfill $0.13625$\hfill
  &&  \hfill $0.13550$\hfill 
  &&  \hfill $1.456(15)$\hfill
  &&  \hfill $0.0539(8)$\hfill
  &&  \hfill $0.143(3)@@$\hfill
  &\cr
\vskip0.3ex
  \+& \hfill $\;$\hfill
  &&  \hfill $$\hfill
  &&  \hfill $0.13590$\hfill 
  &&  \hfill $1.457(19)$\hfill
  &&  \hfill $0.0512(8)$\hfill
  &&  \hfill $0.143(3)@@$\hfill
  &\cr
\vskip0.3ex
  \+& \hfill $\;$\hfill
  &&  \hfill $$\hfill
  &&  \hfill $0.13610$\hfill 
  &&  \hfill $1.464(24)$\hfill
  &&  \hfill $0.0498(8)$\hfill
  &&  \hfill $0.142(3)@@$\hfill
  &\cr
\vskip0.3ex
  \+& \hfill $\;$\hfill
  &&  \hfill $$\hfill
  &&  \hfill $0.13620$\hfill 
  &&  \hfill $1.47(3)@@$\hfill
  &&  \hfill $0.0490(9)$\hfill
  &&  \hfill $0.142(4)@@$\hfill
  &\cr
\vskip0.3ex
  \+& \hfill $\;$\hfill
  &&  \hfill $$\hfill
  &&  \hfill $0.13625$\hfill 
  &&  \hfill $1.48(3)@@$\hfill
  &&  \hfill $0.0487(9)$\hfill
  &&  \hfill $0.142(4)@@$\hfill
  &\cr
\vskip1.5ex
\thicktablerule
\vskip1.5ex
}
$$
\vfill
\endinsert

\beginbibliography


\bibitem{Wilson}
K. G. Wilson,
Phys. Rev. D10 (1974) 2445


\bibitem{Hasenbusch}
M. Hasenbusch,
Phys. Lett. B519 (2001) 177

\bibitem{HasenbuschJansen}
M. Hasenbusch, K. Jansen,
Nucl. Phys. B659 (2003) 299

\bibitem{DellaMorteEtAl}
M. Della Morte et al. (ALPHA collab.),
Comput. Phys. Commun. 156 (2003) 62


\bibitem{SchwarzI}
M. L\"uscher,
JHEP 0305 (2003) 052

\bibitem{SchwarzII}
M. L\"uscher,
Comput. Phys. Commun. 156 (2004) 209

\bibitem{SchwarzIII}
M. L\"uscher,
Comput. Phys. Commun. 165 (2005) 199


\bibitem{UrbachEtAl}
C. Urbach, K. Jansen, A. Shindler, U. Wenger,
Comput. Phys. Commun. 174 (2006) 87


\bibitem{I}
L. Del Debbio, L. Giusti, M. L\"uscher, R. Petronzio, N. Tantalo,
QCD with light Wilson quarks on fine lattices (I): first
experiences and physics results, hep-lat/0610059, to appear in JHEP


\bibitem{GoeckelerEtAl}
M. G\"ockeler et al. (QCDSF--UKQCD collab.),
PoS (LAT2006) 160 and 179


\bibitem{MeyerEtAl}
H. Meyer, O. Witzel (ALPHA collab.),
PoS (LAT2006) 032


\bibitem{IshikawaEtAl}
K-I. Ishikawa et al. (PACS-CS collab.),
PoS (LAT2006) 027

\bibitem{KuramashiEtAl}
Y. Kuramashi et al. (PACS-CS collab.),
PoS (LAT2006) 029

\bibitem{UkawaEtAl}
A. Ukawa et al. (PACS-CS collab.),
PoS (LAT2006) 039


\bibitem{JansenUrbach}
K. Jansen, C. Urbach (ETM collab.),
PoS (LAT2006) 203


\bibitem{SW}
B. Sheikholeslami, R. Wohlert,
Nucl. Phys. B259 (1985) 572

\bibitem{OaImp}
M. L\"uscher, S. Sint, R. Sommer, P. Weisz,
Nucl. Phys. B478 (1996) 365


\bibitem{HMC}
S. Duane, A. D. Kennedy, B. J. Pendleton, D. Roweth,
Phys. Lett. B195 (1987) 216


\bibitem{SextonWeingarten}
J. C. Sexton, D. H. Weingarten,
Nucl. Phys. B380 (1992) 665


\bibitem{Stability}
L. Del Debbio, M. L\"uscher, L. Giusti, R. Petronzio, N. Tantalo,
JHEP 0602 (2006) 011


\bibitem{ZAalpha}
M. Della Morte et al. (ALPHA collab.),
JHEP 0507 (2005) 007


\bibitem{SharpeGap}
S. R. Sharpe,
Phys. Rev. D74 (2006) 014512


\bibitem{GoltermanMobility}
M. Golterman, Y. Shamir, B. Svetitsky,
Phys. Rev. D71 (2005) 071502; {\it ibid.} D72 (2005) 034501


\bibitem{UKQCDlight}
C. R. Allton et al. (UKQCD collab.),
Phys. Rev. D70 (2004) 014501

\bibitem{CPPACSsmall}
Y. Namekawa et al. (CP-PACS collab.),
Phys. Rev. D70 (2004) 074503

\bibitem{ALPHAstability}
M. Della Morte, R. Hoffmann, F. Knechtli, U. Wolff,
Comput. Phys. Commun. 165 (2005) 49


\bibitem{Jacobi}
C. R. Allton et al. (UKQCD collab.),
Phys. Rev. D47 (1993) 5128


\bibitem{NPimp}
K. Jansen, R. Sommer (ALPHA collab.),
Nucl. Phys. B530 (1998) 185
[E: {\it ibid.} B643 (2002) 517]


\bibitem{NPimpCurrent}
M. Della Morte, R. Hoffmann, R. Sommer,
JHEP 0503 (2005) 029


\bibitem{LuscherWeiszTrans}
M. L\"uscher, P. Weisz,
Nucl. Phys. B240 (1984) 349


\bibitem{LuscherResonances}
M. L\"uscher,
Nucl. Phys. B364 (1991) 237

\bibitem{AokiResonances}
S. Aoki et al. (CP-PACS collab.),
PoS (LAT2006) 110


\bibitem{OaImpNonDeg}
T. Bhattacharya, R. Gupta, W. Lee, S. R. Sharpe, J. M. S. Wu,
Phys. Rev. D73 (2006) 034504

\endbibliography

\bye